\documentclass[12pt, a4paper,english,superscriptaddress,aps,nofootinbib]{revtex4}
\usepackage{amsmath,amssymb,graphicx}
\makeatletter
\usepackage{babel}
\usepackage[active]{srcltx}
\usepackage{graphicx,color}
\usepackage{changebar}
\usepackage{hyperref}
\usepackage[T1]{fontenc}
\usepackage{ae}
\usepackage[ansinew]{inputenc}
\usepackage{amsmath}
\usepackage{braket}
\usepackage{mathtools}
\usepackage{slashed}
\usepackage{empheq}
\usepackage{comment}      

\usepackage{graphicx}
\usepackage{amsfonts}
\usepackage{amsmath}
\newcommand{\be}{\begin{equation}}
\newcommand{\ee}{\end{equation}}
\newcommand{\bea}{\begin{eqnarray}}
\newcommand{\eea}{\end{eqnarray}}

\begin{document}

\title{Complex bumblebee model}

\author{Willian Carvalho}
\email[]{willian.carvalho@icen.ufpa.br}
\affiliation{Faculdade de F\'{i}sica, Universidade Federal do Par\'{a}, 66075-110, Bel\'{e}m, Par\'a, Brazil}

\author{A. C. Lehum}
\email[]{lehum@ufpa.br}
\affiliation{Faculdade de F\'{i}sica, Universidade Federal do Par\'{a}, 66075-110, Bel\'{e}m, Par\'a, Brazil}

\author{J. R. Nascimento}
\email[]{jroberto@fisica.ufpb.br}
\affiliation{Departamento de F\'{\i}sica, Universidade Federal da 
Para\'{\i}ba\\
 Caixa Postal 5008, 58051-970, Jo\~ao Pessoa, Para\'{\i}ba, Brazil}
 
\author{A. Yu. Petrov}
\email[]{petrov@fisica.ufpb.br}
\affiliation{Departamento de F\'{\i}sica, Universidade Federal da 
Para\'{\i}ba\\
 Caixa Postal 5008, 58051-970, Jo\~ao Pessoa, Para\'{\i}ba, Brazil}

\begin{abstract}
We formulate a renormalizable complex extension of the bumblebee theory in which the bumblebee field is promoted to a complex one and coupled to an Abelian gauge sector. Besides the minimal gauge covariant interaction, the model includes a longitudinal kinetic term controlled by a dimensionless parameter $g_l$ and a non-minimal magnetic-type coupling $g_m$ between the complex bumblebee and the photon. Using dimensional regularization and minimal subtraction, we determine the one-loop UV divergences of the two-, three-, and four-point functions relevant to the renormalization of the gauge, longitudinal, and quartic sectors. We obtain the corresponding counterterms and derive the one-loop renormalization-group functions for $e$, $g_l$, $g_m$, and the bumblebee self-couplings $\lambda$ and $\tilde\lambda$. Motivated by the known gauge- and field-reparametrization subtleties of the conventional Coleman--Weinberg analysis, we formulate an RG-covariant leading-logarithmic improvement scheme for the Vilkovisky--DeWitt effective potential in normal field coordinates, in which the RG operator is governed solely by the beta functions. We apply this framework to a real constant bumblebee background and obtain the leading-logarithmic one-loop effective potential, discussing the conditions under which a nontrivial vacuum is generated by dimensional transmutation and thereby provides a dynamical realization of Lorentz symmetry breaking in this class of models.
\end{abstract}

\maketitle

\section{Introduction}

Spontaneous Lorentz symmetry breaking (LSB) is actually treated as the most natural way to implement Lorentz symmetry violation within the curved space-time context where, in general, constant Lorentz-violating (LV) vectors cannot be consistently defined, see the discussion in \cite{KosGra}. Conceptually, within this scenario a LV vector is not introduced  {\it a priori}, as within the explicit Lorentz-breaking scenario, but arises dynamically. Within the bumblebee model, originally introduced in \cite{KosGra} and providing a most efficient framework to break the Lorentz symmetry spontaneously, such a vector arises as one of minima of the potential of some vector field called the bumblebee field.  Further, various issues related to the bumblebee model have been considered, in particular, studies of its coupling to gravity (for a general review on LV gravity see \cite{KosLiGrav}) within the black hole context \cite{Bertolami,sol2,L}, cosmological context \cite{Capelo,H}, weak gravity context \cite{Maluf1}, and metric-affine gravity \cite{ourbumb1,ourbumb2} have been performed. 

At the same time, studies of the bumblebee model in a flat space-time also present a certain interest, allowing for dynamical generating the known LV models. The most interesting direction of study in this case is the dynamical LSB where the bumblebee potential is generated as a quantum correction \cite{dyn1,dyn2,dyn3}. Another direction is the study of interaction of the bumblebee field with other fields. Some examples of couplings of this field to scalar, spinor and gauge fields can be found in \cite{ourbumb2,ourbumb3,Lehum:2024ovo}, where such a coupling were motivated by the metric-affine bumblebee gravity originally introduced in \cite{ourbumb1}.  

In this paper we formulate a renormalizable complex extension of the bumblebee model coupled to an Abelian gauge field. Our construction differs from the previously studied real bumblebee setups in several important respects. First, promoting the bumblebee field to a complex one allows for two independent renormalizable quartic invariants in the self-interaction sector, rather than a single quartic structure. Second, we carry out the complete one-loop renormalization of the coupled Abelian theory, determining the counterterms and beta functions for the gauge, longitudinal, magnetic-type, and quartic sectors. Third, motivated by the gauge- and parametrization-dependence of the conventional Coleman--Weinberg (CW) analysis, we formulate an RG-covariant leading-logarithmic (LL) treatment of the Vilkovisky--DeWitt (VDW) effective potential and apply it to the real constant-background sector of the model. Together, these results provide a unified setting in which the richer self-coupling structure of the complex theory, its renormalization properties, and the possibility of radiatively induced LSB can be analyzed on the same footing.

The paper is organized as follows. In Sec.~II we define the complex bumblebee model coupled to an Abelian gauge field and present the renormalized and counterterm Lagrangians that set the stage for perturbation theory. In Sec.~III we compute the one-loop UV divergences in dimensional regularization and extract the minimal-subtraction counterterms and beta functions for all couplings of the model. In Sec.~IV we outline an RG improvement scheme for the VDW effective potential formulated in RG-covariant (normal) field coordinates, emphasizing the resulting RG equation controlled solely by beta functions. In Sec.~V we apply this formalism to the real constant bumblebee background, derive the LL one-loop effective potential, and analyze the emergence of a radiatively induced vacuum through dimensional transmutation. Our conclusions and outlook are collected in Sec.~VI.

\section{The massless complex bumblebee model}

Let us define the model with the following renormalizable Lagrangian density: 
\begin{eqnarray}\label{eq001}
\mathcal{L} &=& 
-\frac{1}{4} F_{\mu\nu} F^{\mu\nu}
-\frac{1}{2} (D_{\mu} B_{\nu} - D_{\nu} B_{\mu}) (D^{\mu} B^{\nu} - D^{\nu} B^{\mu})^*
-\frac{1}{g_l} (D^\mu B_\mu) (D^\mu B_\mu)^* \nonumber \\
&&
- i g_{m} B^{*\mu} B^{\nu} F_{\mu\nu}
-\frac{\lambda}{4} (B^{*\mu} B_{\mu})^2
-\frac{\tilde{\lambda}}{4} (B^{*\mu} B^{\nu} B^*_{\mu} B_{\nu})
+ \mathcal{L}_{\mathrm{CT}},
\end{eqnarray}
where $D_{\mu} = \partial_{\mu} - i e A_{\mu}$ is the usual covariant derivative, $F_{\mu\nu}$ is the field strength tensor associated with the electromagnetic field $A_{\mu}$, and $\mathcal{L}_{\mathrm{CT}}$ denotes the counterterm Lagrangian. The parameters $g_l$ and $g_m$ respectively characterize the coupling associated with the longitudinal kinetic term of the bumblebee field and the bumblebee non-minimal magnetic-type interaction with the electromagnetic field.

In contrast to the usual real bumblebee setup, the complex bumblebee allows for two independent renormalizable quartic invariants. Besides the ``single-trace'' structure built from the scalar contraction $(B^{*\mu}B_\mu)$, one can also write an inequivalent quartic contraction in which Lorentz indices are paired across the complex-conjugate fields. The most general quartic self-interaction consistent with Lorentz symmetry and the $U(1)$ phase symmetry $B_\mu\to e^{i\alpha}B_\mu$ therefore reads
\begin{equation}
V(B,B^*)
=
\frac{\lambda}{4}\,\big(B^{*\mu}B_\mu\big)^2
+\frac{\tilde{\lambda}}{4}\,\big(B^{*\mu}B^{\nu}B^*_{\mu}B_{\nu}\big),
\label{eq:complex_quartic}
\end{equation}
where $\lambda$ and $\tilde\lambda$ are \textit{a priori} independent dimensionless couplings. This feature is absent in the real bumblebee case, for which only a single quartic invariant is available. As a result, the complex theory exhibits a genuinely richer renormalizable self-coupling sector, with two independent RG flows that can affect the structure of radiative corrections and the vacuum configuration.

The explicit form of the counterterm Lagrangian is given by
\begin{eqnarray}
\mathcal{L}_{\mathrm{CT}} &=& 
- \frac{\delta_3}{4} F_{\mu\nu} F^{\mu\nu}
- \frac{\delta_2}{2} B_{\mu\nu} B^{*\mu\nu}
- e^2 \delta_4
\left(
A_{\mu} A^{\mu} B_{\nu} B^{*\nu}
-
A_{\mu} A_{\nu} B^{\mu} B^{*\nu}
\right)
\nonumber \\
&&
- \frac{i e \, \delta_1}{2}
\left[
B_{\mu\nu} (A^{\mu} B^{*\nu} - A^{\nu} B^{*\mu})
-
B^*_{\mu\nu} (A^{\mu} B^{\nu} - A^{\nu} B^{\mu})
\right]  \nonumber \\
&&- \frac{\delta_{g_l}}{g_l} \partial^\mu B^*_\mu \partial^\nu B_\nu
+ i e \frac{\delta_5}{g_l}
\left(
A^\mu B_{\mu} \partial^\nu B^*_\nu
-
A^\mu B^*_{\mu} \partial^\nu B_\nu
\right) 
- e^2 \frac{\delta_6}{g_l} A^\mu A^\nu B^*_\mu B_\nu
\nonumber \\
&&
- i \delta_{g_m} B^{*\mu} B^{\nu} F_{\mu\nu} 
- \frac{\delta_\lambda}{4} (B^{*\mu} B_\mu)^2
- \frac{\delta_{\tilde{\lambda}}}{4} (B^{*\mu} B^{\nu} B^*_{\mu} B_{\nu}).
\end{eqnarray}
\noindent In the expressions above, all fields and couplings are to be understood as renormalized quantities.

The counterterm Lagrangian is obtained by rewriting the bare Lagrangian in terms of renormalized fields and couplings. The relations between the bare quantities (denoted by the subscript $0$) and the renormalized ones are established through the introduction of the wave-function renormalization constants,
\begin{eqnarray}
{B_0}_\mu \rightarrow Z_2^{1/2} B_\mu, 
\quad {B_0}^*_\mu \rightarrow Z_2^{1/2} B^*_\mu, 
\quad {A_0}_\mu \rightarrow Z_3^{1/2} A_\mu,
\end{eqnarray}
from which the renormalized couplings are defined as
\begin{eqnarray}
&& Z_2 \equiv 1 + \delta_2, \hspace{1cm} Z_3 \equiv 1 + \delta_3, \nonumber \\
&& e_0 Z_3^{1/2} Z_2 \equiv e (1 + \delta_1), \hspace{1cm} e_0^2 Z_3 Z_2 \equiv e^2 (1 + \delta_4), \nonumber \\
&& \lambda_0 Z_2^2 \equiv \lambda + \delta_\lambda, \hspace{1cm} \tilde{\lambda}_0 Z_2^2 \equiv \tilde{\lambda} + \delta_{\tilde{\lambda}}, \nonumber \\
&& g_{m0} Z_3^{1/2} Z_2 \equiv g_m + \delta_{g_m}, \hspace{1cm} \frac{Z_2}{g_{l0}} = \frac{1 + \delta_{g_l}}{g_l}, \nonumber \\
&& \frac{e_0 Z_3^{1/2} Z_2}{g_{l0}} = \frac{e (1 + \delta_5)}{g_l}, \hspace{1cm} \frac{e_0^2 Z_3 Z_2}{g_{l0}} = \frac{e^2 (1 + \delta_6)}{g_l}.
\end{eqnarray}
These relations will be employed in the computation of the renormalization group functions.

It is important to note that the inclusion of the longitudinal term for the bumblebee field leads to the following form for the propagator:
\begin{eqnarray}
\langle B^{\mu}(p) B^{*\nu}(-p) \rangle &=& \frac{i}{p^2} \left( T^{\mu\nu}(k) + g_l L^{\mu\nu}(p) \right),
\end{eqnarray}
where the transverse and longitudinal projectors are defined, respectively, as
\begin{eqnarray}
T^{\mu\nu}(p) &=& \eta^{\mu\nu} - \frac{p^{\mu} p^{\nu}}{p^2}~\hspace{.5cm} \mathrm{and} \hspace{.5cm}
L^{\mu\nu}(p) = \frac{p^{\mu} p^{\nu}}{p^2}.
\end{eqnarray}

It is important to emphasize that the inclusion of the longitudinal term in the bumblebee field propagator requires careful consideration to ensure the consistency of the quantization procedure. In particular, the absence of ghost states and the preservation of unitarity demand that the norm associated with the longitudinal component of the propagator be positive definite. Consequently, to avoid the emergence of negative-norm states in the Hilbert space, the coupling parameter $g_l$, which controls the longitudinal contribution, must satisfy the condition $g_l > 0$. Under this condition, the longitudinal mode contributes consistently to the dynamics of the theory without violating unitarity. In contrast, for $g_l < 0$, the longitudinal sector would exhibit negative-norm states, leading to unphysical ghost excitations that may render the theory inconsistent.

\section{RENORMALIZATION GROUP FUNCTIONS}

\begin{figure}[ht!]
	\includegraphics[angle=0 ,width=15cm]{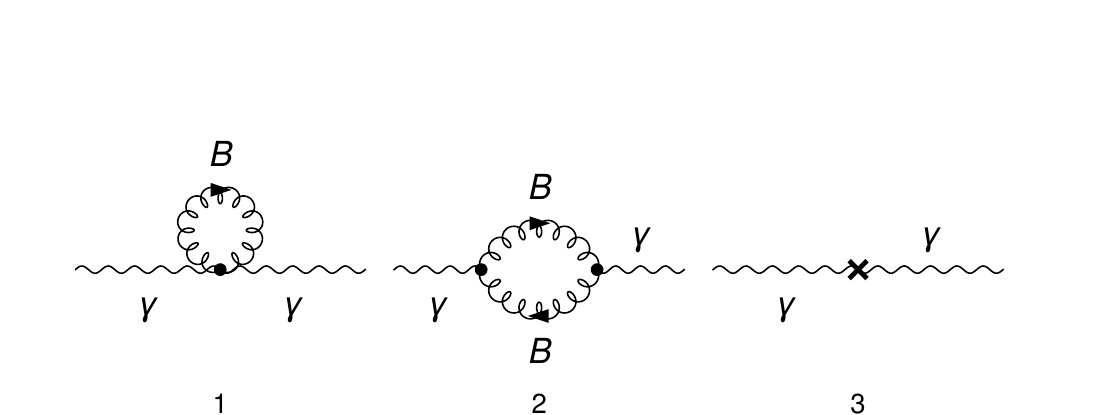}
	\caption{One-loop bumblebee contributions to the photon two-point function (vacuum polarization). Wavy and wiggly lines denote the photon and bumblebee propagators, respectively.}
	\label{fig01}
\end{figure}

In this section, we present the UV renormalization of the complex bumblebee model. We begin with the
bumblebee corrections to the photon propagator. The corresponding Feynman diagrams are shown in Figure~\ref{fig01}. In order to compute the Feynman diagrams we used a set of Mathematica\textsuperscript{TM} packages \cite{feyncalc,feyncalc1,feyncalc2,feynarts,feynrules,feynrules1,feynhelpers}. 

The  expression corresponding to the diagram \ref{fig01}.1 is given by
\begin{eqnarray}
\Pi^{\mu\nu}_1 &=&  \frac{ e^2}{g_l} \int \frac{d^4k}{(2\pi)^4}
\left[
(g_l-1) (\eta^{\alpha \mu} \eta^{\gamma \nu}
+ \eta^{\alpha \nu} \eta^{\gamma \mu})
-2 g_l \eta^{\alpha \gamma} \eta^{\mu \nu}
\right]
\left[
\frac{\eta_{\alpha \gamma}}{k^2}
+\frac{(g_l-1) k_{\alpha} k_{\gamma}}{k^4}
\right].
\end{eqnarray}
We observe that the amplitude above vanishes identically, as $\textbf{A}_0(0)=0$, as is expected for massless fields.

For the diagram shown in Figure~\ref{fig01}.2, we obtain the corresponding expression
\begin{eqnarray}
\Pi^{\mu\nu}_2 &=& \frac{1}{g_l^2}\int \frac{d^4k}{(2\pi)^4}\Bigg[
\eta^{\beta \nu} \big(e \left(g_l (p-k)^{\alpha} + k^{\alpha} \right) + g_l g_m p^{\alpha} \big)
-\eta^{\alpha \nu} \big(e \left(g_l k^{\beta} + (p-k)^{\beta} \right)  \nonumber\\
&& \quad+ g_l g_m p^{\beta} \big)
+ e g_l \eta^{\alpha \beta} \left(k^{\nu} - (p-k)^{\nu} \right)\Bigg]\times \Bigg[
\eta^{\delta \mu} \left( e \left( k^{\gamma} - g_l (k - p)^{\gamma} \right) + g_l g_m p^{\gamma} \right)\nonumber\\
&&\quad- \eta^{\gamma \mu} \left( e g_l k^{\delta} - e (k - p)^{\delta} + g_l g_m p^{\delta} \right)
+ e g_l \eta^{\gamma \delta} \left( k^{\mu} + (k - p)^{\mu} \right)\Bigg]\times \Bigg[
\eta_{\beta \delta} \bigg(
\frac{\eta_{\alpha \gamma}}{k^2 (k - p)^2} \nonumber\\
&& \quad+ \frac{(g_l - 1) k_{\alpha} k_{\gamma}}{k^4 (k - p)^2}
\bigg)
- (g_l - 1) (p - k)_{\beta} (k - p)_{\delta} \left(
\frac{\eta_{\alpha \gamma}}{k^2 (k - p)^4}
+ \frac{(g_l - 1) k_{\alpha} k_{\gamma}}{k^4 (k - p)^4}
\right)\Bigg]\nonumber \\ \nonumber
&=& \frac{i \pi^2 }{12}\left(p^{\mu} p^{\nu} - p^2 \eta^{\mu\nu}\right)\Bigg[
(g_l - 1) \Big(2  \left(2 e (e + 2 g_m) + (g_l + 1) g_m^2\right)\textbf{B}_0(0, 0, 0)\nonumber\\
&& \quad-4 p^2 \big(e^2 + 2 e g_m + g_l g_m^2\big) \textbf{C}_0(0, p^2, p^2, 0, 0, 0)
\nonumber\\
&& \quad+ (g_l - 1) g_m^2 p^4 \textbf{D}_0(0, p^2, 0, p^2, p^2, p^2, 0, 0, 0, 0)\Big)
\nonumber\\
&& \quad+ 2 \left(4 e^2 (g_l - 2) - 4 e (g_l + 5) g_m - (g_l + 1)(g_l + 5) g_m^2
\right) \textbf{B}_0(p^2, 0, 0)\Bigg],
\end{eqnarray}
where the scalar integrals $\textbf{B}_0$, $\textbf{C}_0$ and $\textbf{D}_0$ follow the Passarino-Veltman prescription and are defined in the Appendix A. 

After evaluating the integrals, we obtain the relation for the amplitude, which contains both finite and UV-divergent contributions, and is given by
\begin{eqnarray}
    i\Pi^{\mu\nu} &=& \left(p^{\mu} p^{\nu} - p^2 g^{\mu\nu}\right)\Bigg\{ \frac{1}{48 \pi^2 \epsilon} \left(e^2 (5 - 3 g_l) + 12 e g_m + 3 (g_l + 1) g_m^2\right)   \nonumber\\
&&-\frac{1}{288 \pi^2}\Big[2 e^2 \Big(9 g_l(1- \gamma) + 3 (3 g_l - 5) \ln\left(\frac{-4 \pi \mu^2 }{p^2}  \right) + 15 \gamma - 19 
\Big)\nonumber\\
&&-36 e g_m \Big(g_l + 2\ln\left(\frac{-4 \pi \mu^2 }{p^2}  \right) - 2 \gamma + 3\Big)\nonumber\\
&&-9 (g_l + 1) g_m^2 \Big( g_l+2\ln\left(\frac{-4 \pi \mu^2 }{p^2}  \right)
   - 2\gamma + 3 \Big)\Big]\Bigg\},
\end{eqnarray}
where $\epsilon=(4-D)/2$. We employ dimensional regularization (DR) and the minimal subtraction (MS) scheme as our renormalization prescriptions. By including the counterterm diagram shown in Figure~\ref{fig01}.3 and imposing the finiteness condition, we obtain the corresponding counterterm
\begin{eqnarray}
    \delta_3 &=& \frac{1}{48 \pi^2 \epsilon}\left(e^2 (5 - 3 g_l) + 12 e\, g_m + 3 (g_l + 1) g_m^2\right).
\end{eqnarray}

\begin{figure}[ht!]
	\includegraphics[angle=0,width=15cm]{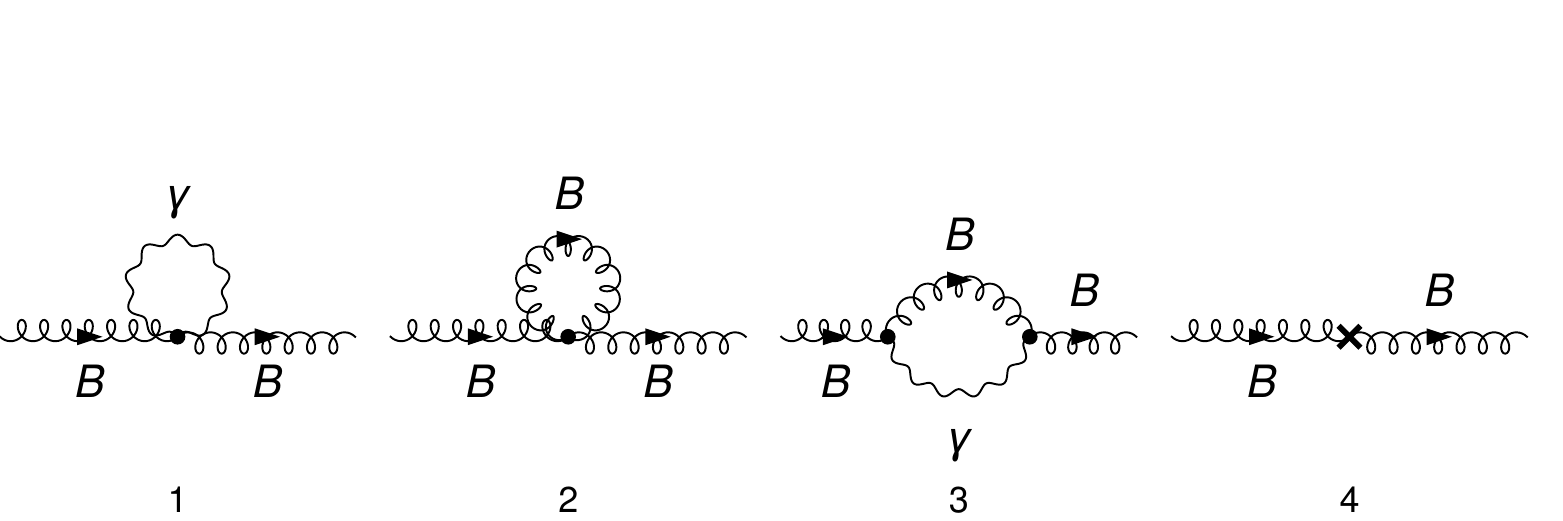}
	\caption{Bumblebee self-energy (two-point 1PI function).}
	\label{fig02}
\end{figure}

We now compute the self-energy of the bumblebee field corresponding to the Feynman diagram shown in Figure~\ref{fig02}. For the first and second diagrams, the corresponding loop integrals are given by
\begin{subequations}
\begin{eqnarray}
\Gamma_1^{\mu\nu} &=& \frac{e^2}{2 g_l} \!\int\! \frac{d^4k}{(2\pi)^4}
\left[
\frac{\eta_{\alpha \gamma}}{k^2}
+ \frac{k_\alpha k_\gamma (\xi - 1)}{k^4}
\right]
\left[
(g_l - 1) \left(\eta^{\alpha \mu} \eta^{\gamma \nu} + \eta^{\alpha \nu} \eta^{\gamma \mu} \right)
- 2 g_l \, \eta^{\alpha \gamma} \eta^{\mu \nu}
\right]; \\
\Gamma_2^{\mu\nu} &=& \frac{1}{2} \int \frac{d^4k}{(2\pi)^4}
\left[
\frac{\eta_{\alpha \gamma}}{k^2}
+ \frac{(g_l - 1) \, k_{\alpha} k_{\gamma}}{k^4}
\right]
\left[
2\tilde{\lambda} \, \eta^{\alpha \nu} \eta^{\gamma \mu}
- \lambda \left( \eta^{\alpha \mu} \eta^{\gamma \nu} + \eta^{\alpha \gamma} \eta^{\mu \nu} \right)
\right].
\end{eqnarray}
\end{subequations}
Analogously to the previous case, we observe that the integrals are proportional to $\textbf{A}_0(0)$ and therefore vanish identically, as they involve massless fields.

For the third diagram, Figure~\ref{fig02}.3, we have
\begin{eqnarray}
\Gamma_3^{\mu\nu} &=&  \frac{1}{g_l^2} \int \frac{d^4k}{(2\pi)^4} \Bigg[
e \Big(g_l \eta^{\beta \nu} \left((p-k)^{\alpha}+p^{\alpha}\right)
- g_l \eta^{\alpha \beta} (p-k)^{\nu}
+ p^{\nu} \eta^{\alpha \beta}\Big)
\nonumber\\
&&-(\eta^{\alpha \nu} \big(e \left(g_l p^{\beta}-(p-k)^{\beta}\right)+g_l g_m k^{\beta}\big))
+ g_l g_m k^{\nu} \eta^{\alpha \beta}
\Bigg]  \times
\Bigg[
g_l g_m k^{\mu} \eta^{\gamma \delta}
\nonumber\\
&&-(\eta^{\gamma \mu} \big(e \left(g_l p^{\delta}+(k-p)^{\delta}\right)+g_l g_m k^{\delta}\big))
+ e \Big(g_l \eta^{\delta \mu} \left(p^{\gamma}-(k-p)^{\gamma}\right)
\nonumber\\
&&+ g_l \eta^{\gamma \delta} (k-p)^{\mu}
+ p^{\mu} \eta^{\gamma \delta}\Big)
\Bigg]
\times
\Bigg[
\frac{k_{\alpha} k_{\gamma}  \eta_{\beta \delta}}{k^4 (k-p)^2}(\xi-1)
+\frac{\eta_{\alpha \gamma} \eta_{\beta \delta}}{k^2 (k-p)^2}
\nonumber\\
&&
-(g_l-1) (p-k)_{\beta} (k-p)_{\delta}
\left(
\frac{\eta_{\alpha \gamma}}{k^2 (k-p)^4}
+\frac{k_{\alpha} k_{\gamma} }{k^4 (k-p)^4}(\xi-1)
\right)
\Bigg].
\end{eqnarray}

\begin{figure}[h!]
	\includegraphics[angle=0,width=15cm]{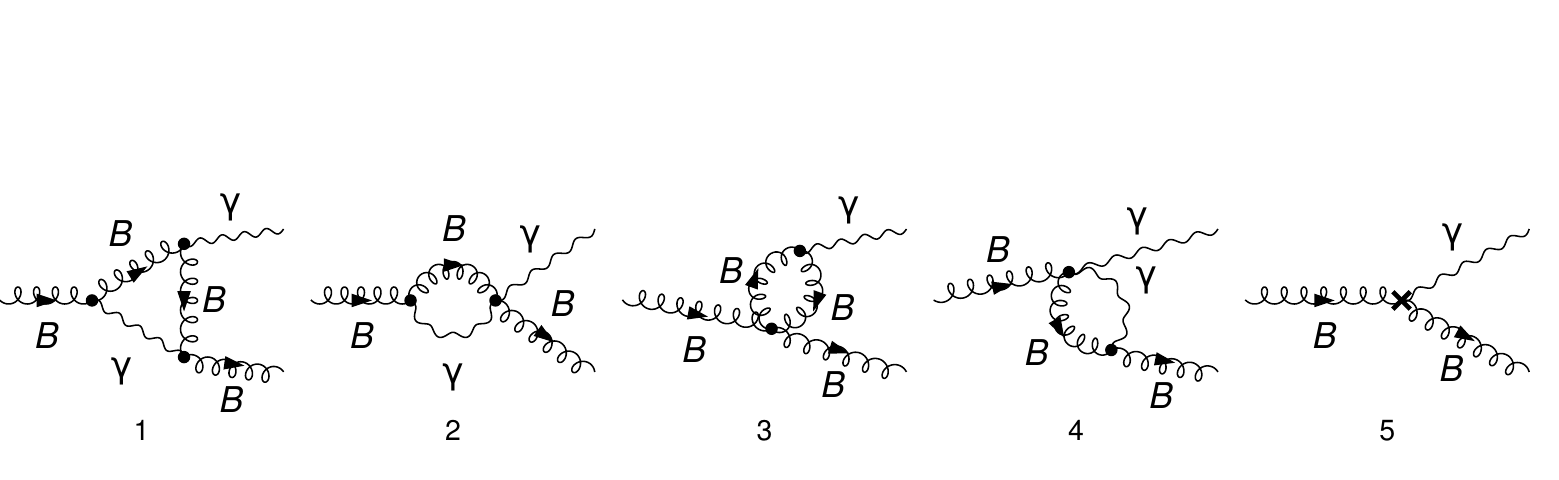}
	\caption{The bumblebee--bumblebee--photon three-point function (1PI vertex).}
	\label{fig03}
\end{figure}

After evaluating the integral and adding the counterterm contribution from Figure~\ref{fig02}.4, we obtain
\begin{eqnarray}
    i\Gamma^{\mu\nu} &=& p^2\left[T^{\mu\nu}\left(\frac{\delta_2}{g_m} - \Gamma_T\right) + L^{\mu\nu}\left(\frac{\delta_{g_l}}{g_l}- \Gamma_L\right)\right],
\end{eqnarray}
where
\begin{subequations}
\begin{eqnarray}
    \Gamma_T &=& \frac{1}{96 \pi^2 g_m\epsilon}\left(e^2 \left(3 g_l - 6 \xi + 14\right) + e \left(g_m - 3 g_l g_m\right) + (2 - 3 g_l) g_m^2\right)\nonumber  \\  
    &&-\frac{1}{288 \pi^2 g_m}\Big[e^2\Big(3 \left(3 g_l - 6 \xi + 14\right)\ln\bigg(\frac{-4 \pi \mu^2 }{p^2}  \bigg)+9 g_l \xi- 9 \gamma g_l + 18 g_l \nonumber\\  &&+\xi (18 \gamma - 9)+49-42 \gamma\Big)+e g_m\Big((3-9g_l)\ln\bigg(\frac{-4 \pi \mu^2 }{p^2}  \bigg)+9 \gamma  g_l-3\gamma-10\Big) \nonumber\\ 
    &&+g_m^2\Big((6-9g_l)\ln\bigg(\frac{-4 \pi \mu^2 }{p^2}  \bigg)+9 \gamma  g_l-9 g_l-6 \gamma+4\Big)\Big];\\
    \Gamma_L &=& \frac{1}{32 \pi^2 g_l^2\epsilon}\left(e^2 \left(g_l^2 - 2 g_l \, \xi + 6\right) - e (g_l - 3) g_l g_m + g_l^2 g_m^2\right) \nonumber \\  
    &&-\frac{1}{32 \pi ^2 g_l^2}\Bigg[e^2\Big((g_l^2 + 6)\ln \bigg(\frac{-4 \pi \mu^2 }{p^2}  \bigg)- 2 \xi \Big( g_l \ln\bigg(\frac{- \pi \mu^2 }{p^2}  \bigg) + \ln(4) \Big)\Big) \nonumber \\  
    && + e g_l g_m \Big((3-g_l)\Big(\ln \bigg(\frac{-4 \pi \mu^2 }{p^2}  \bigg)-\gamma\Big)  -2 g_l+4\Big)\nonumber\\
    && +g_l^2 g_m^2\Big(\ln\bigg(\frac{- \pi \mu^2 }{p^2}  \bigg)-\gamma+1\Big)\Bigg].
\end{eqnarray}
\end{subequations}
By imposing finiteness through the MS scheme, we find the counterterms to be $\delta_2=g_m \Gamma_T $ and $\delta_{g_l}={g_l}\Gamma_L$. The divergent part of these counterterms can be written as
\begin{subequations}
\begin{eqnarray}\label{eq:delta2}
    \delta_2 &=& \frac{1}{96 \pi^2 \epsilon}\Big[e^2 \left(3 g_l - 6 \xi + 14\right) + e \left(g_m - 3 g_l g_m\right) + (2 - 3 g_l) g_m^2\Big];\\ 
    \delta_{g_{l}} &=& \frac{1}{32 \pi^2 g_l\epsilon}\Big[e^2 \left(g_l^2 - 2 g_l  \xi + 6\right) - e (g_l - 3) g_l g_m + g_l^2 g_m^2\Big].
\end{eqnarray}
\end{subequations}
In the following, we compute the counterterms corresponding to renormalization of the couplings $e$, $e/g_l$ and $g_m$. The relevant Feynman diagrams required to determine the corresponding renormalization constants are shown in Figure~\ref{fig03}. The UV-divergent part of the associated three-point function is given by 
\begin{eqnarray}
\langle T B^{\mu}B^{\nu}A^{\gamma}\rangle &=& \eta^{\nu\gamma} p^\mu \bigg[\frac{i}{192 \pi^2 g_l^2 \epsilon}\Big(-2 e^3 \left(7 g_l^2 + 6 g_l \xi - 18\right)
- 2 e^2 g_l g_m \big(6 g_l^2 + 6 g_l \xi- 4 g_l  - 9\big)\nonumber \\ \nonumber 
&& 
- 2 e g_l^2 \big( (3 g_l + 4) g_m^2 - 3 \lambda - 6 \tilde{\lambda}\big)
+ 3 g_l^2 g_m \big(   (g_l+1)(\lambda + 2\tilde{\lambda})-2 g_l g_m^2  \big)\Big) \\ \nonumber 
&& - \frac{i }{g_l}\left( \delta_5  e + \delta_{g_m} \, g_l \right)\bigg]+\eta^{\mu\gamma}p^\nu \bigg[\frac{ie }{96 \pi^2 \epsilon} \Big( 
e^2 (3 g_l - 6 \xi + 14) + e (g_m - 3 g_l g_m) \\ \nonumber 
&&+ (2 - 3 g_l) g_m^2 \Big)- i \delta_1 e\bigg]
+\eta^{\mu\nu}p^\gamma \bigg[i \left( \delta_{g_m} + \delta_1 \, e \right)- \frac{i}{192 \pi^2 \epsilon}\Big(
2 e^3 (3 g_l - 6 \xi + 4)\\ \nonumber 
&&- 2 e^2 g_m \left(9 g_l + 6 \xi - 8\right)- 2 e \big(  (6 g_l +5) g_m^2 - 3 \lambda - 6 \tilde{\lambda} \big)+ 3 g_m ( -2 g_l g_m^2+ \lambda \\ 
&&+ g_l \lambda + 2 g_l \tilde{\lambda}  + 2 \tilde{\lambda} )\Big)\bigg].
\end{eqnarray}

Thus, using the MS renormalization scheme, the counterterms $\delta_1$, $\delta_5$ and $\delta_{g_m}$ are readily determined from the divergent part of the corresponding three-point functions. They are given by
\begin{subequations}
    \begin{eqnarray}
    \delta_1 &=& \frac{1}{96 \pi^2 \epsilon}\Big[e^2 \left(3 g_l - 6 \xi + 14\right)+ e \left(g_m - 3 g_l g_m\right)+ \left(2 - 3 g_l\right) g_m^2\Big];\label{eq:delta1} \\
    \delta_5 &=& \frac{1}{32 \pi^2 g_l\epsilon}\Big[e^2 \left(g_l^2 - 2 g_l \xi+ 6\right)- e \left(g_l - 3\right) g_l g_m+ g_l^2 g_m^2\Big];\\
    \delta_{g_m} &=& -\frac{1}{
192 \pi^2 \epsilon
}\Big[20 e^3
+ 2 e^2 g_m (6 g_l + 6 \xi - 7)
- 2 e \big( 3\lambda + 6\tilde{\lambda} - (3 g_l + 7) g_m^2  \big)
\nonumber \\ 
&&- 3 g_m \big( (\lambda + 2\tilde{\lambda})(g_l+1)-2 g_l g_m^2  \big)
\Big].
\end{eqnarray}
\end{subequations}

\begin{figure}[ht!]
	\includegraphics[angle=0,width=15cm]{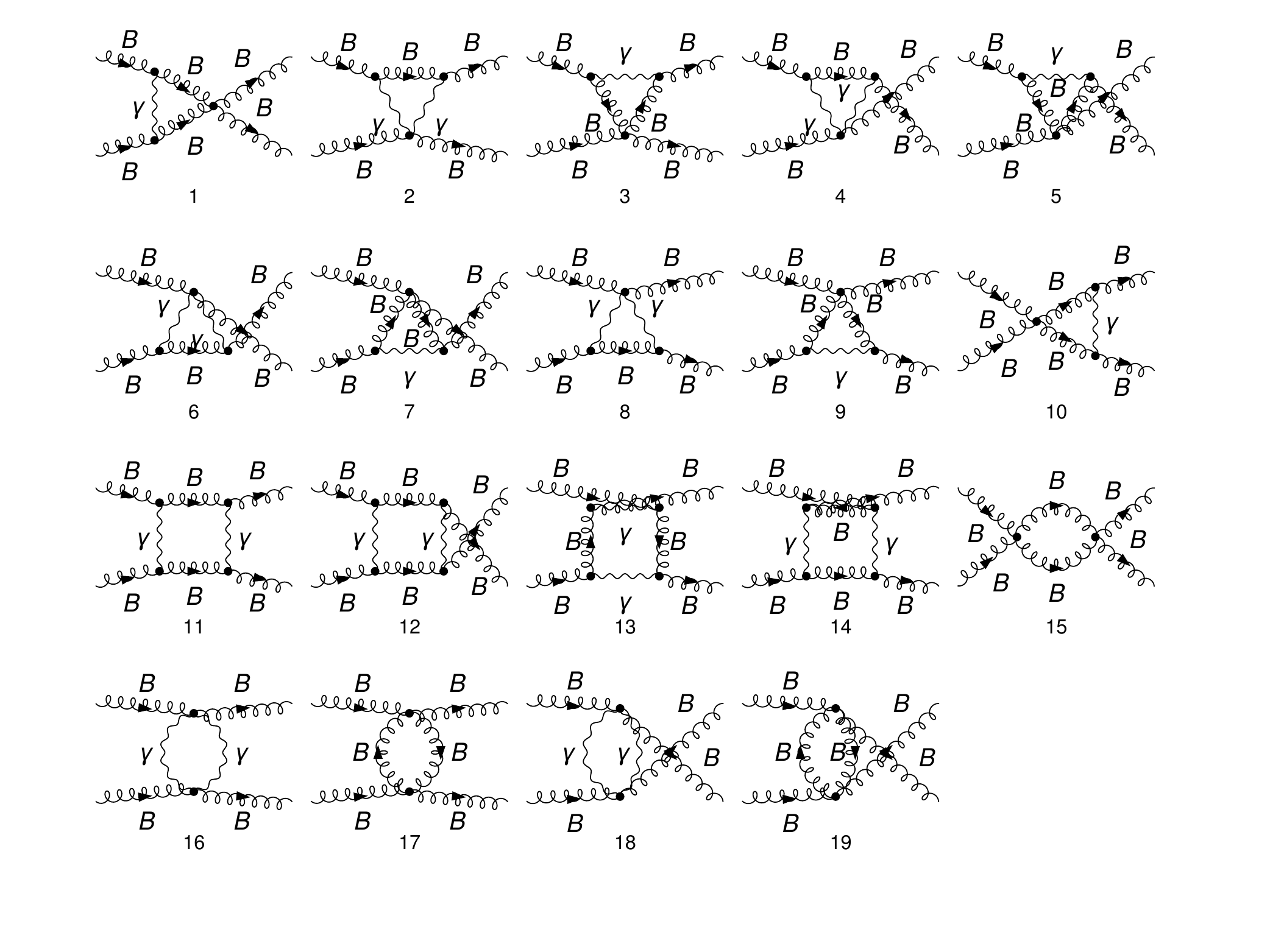}
	\caption{One-loop 1PI bumblebee four-point function. Its UV divergence determines the counterterms required to renormalize the quartic self-interactions governed by $\lambda$ and $\tilde{\lambda}$.}
	\label{fig04}
\end{figure}

Finally, we proceed with the renormalization of the four-point function of the bumblebee field, as represented by the diagrams in Figure~\ref{fig04}. Our calculations were performed in the limit where the external momenta of the bumblebee field legs are set to zero, which is sufficient to study the effective potential. The associated UV divergent part is expressed as
\begin{eqnarray}
\langle T B^{\mu}B^{\nu}B^{\alpha}B^{\gamma}\rangle =
(\eta^{\nu\alpha}\eta^{\mu\gamma}
+\eta^{\mu\alpha}\eta^{\nu\gamma})\Gamma^{(4)}_1 
+\eta^{\mu\nu}\eta^{\alpha\gamma}\Gamma^{(4)}_2,
\end{eqnarray}
\noindent where  
\begin{subequations}
    \begin{eqnarray}\label{fourpoints}
\Gamma^{(4)}_1 &=& - \frac{i\,\delta \lambda}{2}+\frac{i}{192 \pi^2\epsilon}\Big[ 22 e^4 - 56 e^3 g_m - 6 e^2 \left(2 (g_l - 6) g_m^2 + 5 \lambda + 2 \lambda \xi \right)  \nonumber \\ \nonumber 
&&+ e \big(8 (6 g_l + 5) g_m^3- 30 (g_l + 1) g_m \lambda \big) + 8 g_l^2 g_m^4 - 2 \lambda \big(3 (2 g_l^2 + g_l + 2) g_m^2  \\ 
&& + (2 g_l^2 - g_l + 5) \tilde{\lambda} \big)+ 4 g_l^2 g_m^2 \tilde{\lambda}+ (5 g_l^2 + 2 g_l + 17) \lambda^2 + g_l^2 \tilde{\lambda}^2 
+ 8 g_l g_m^4 \nonumber\\ 
&& - 8 g_l g_m^2 \tilde{\lambda} + 4 g_l \tilde{\lambda}^2 
+ 6 g_m^4 + 4 g_m^2 \tilde{\lambda} + 7 \tilde{\lambda}^2\Big]
 ;\\ \nonumber\label{fourpoints2}
\Gamma^{(4)}_2 &=& i \delta \tilde{\lambda} 
+ \frac{i}{384 \pi^2 \epsilon}\Big[ 32 e^4 + 128 e^3 g_m + 24 e^2 \big( g_m^2(2 g_l + 6 ) + 2 \lambda -  \tilde{\lambda}(3- 2\xi)    \big) \\ \nonumber 
&& + 8 e g_m \big(4 (3 g_l + 1) g_m^2 + 6 (g_l + 1) \lambda - 9 (g_l + 1) \tilde{\lambda} \big) + 4 (7 g_l^2 - 2 g_l + 3) g_m^4 \\ \nonumber 
&&+ (g_l - 1)^2 \lambda^2- 8 (5 g_l^2 - g_l + 5) g_m^2 \tilde{\lambda} - 4 (2 g_l^2 + 5 g_l + 11) \lambda \tilde{\lambda} 
+ 12 (g_l + 1)^2 g_m^2 \lambda\\ 
&& + 2 (7 g_l^2 - 2 g_l + 19) \tilde{\lambda}^2\Big],
\end{eqnarray}
\end{subequations}
with the first two terms arise from the counterterm diagrams and can be determined by imposing finiteness conditions on the expressions \eqref{fourpoints} and \eqref{fourpoints2}. 

\begin{figure}[ht!]
	\includegraphics[angle=0,width=15cm]{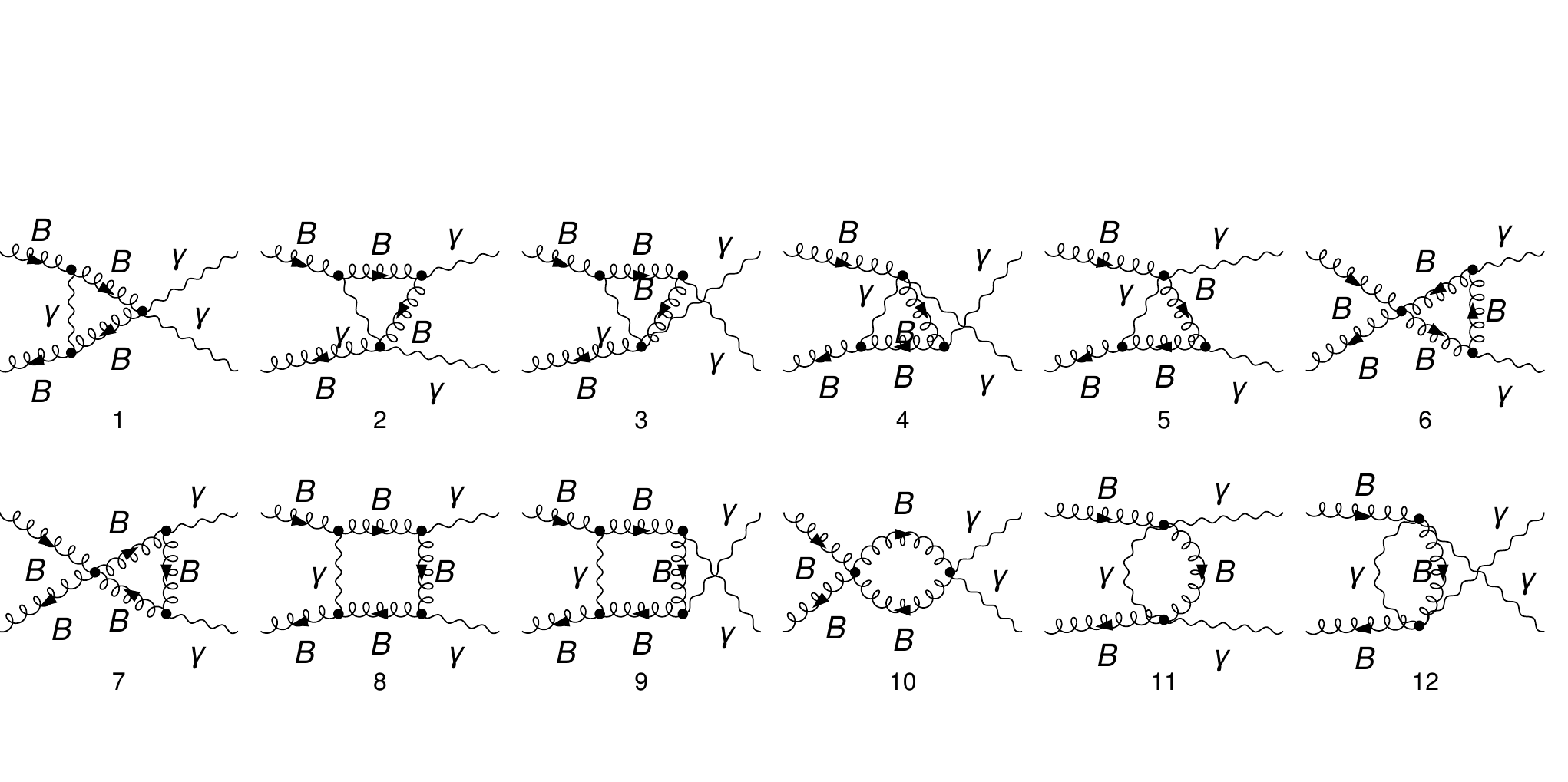}
	\caption{One-loop 1PI contribution to the bumblebee--bumblebee--photon--photon four-point function.}
	\label{fig05}
\end{figure}

We also calculate the one-loop four-point function $\langle T B^{\mu}B^{\nu}A^{\alpha}A^{\gamma}\rangle$ to find the counterterms $\delta_4$ and $\delta_6$ represented in the diagrams of the Figure~\ref{fig05}. The expression to the divergent part is
\begin{eqnarray}
\langle T B^{\mu}B^{\nu}A^{\alpha}A^{\gamma}\rangle =
(\eta^{\mu\alpha}\eta^{\nu\gamma}
+\eta^{\nu\alpha}\eta^{\mu\gamma})\Gamma^{(4)}_3 
+\eta^{\mu\nu}\eta^{\gamma\alpha}\Gamma^{(4)}_4,
\end{eqnarray}
\noindent where  
\begin{subequations}
\begin{eqnarray}\label{fourpointsBA}
\Gamma^{(4)}_3 &=& \frac{i e^2}{g_l} \left( \delta_4 g_l - \delta_6 \right)+\frac{i e^2 }{96 \pi^2 g_l^2 \epsilon}\Big[ 
e^2 \left( 6 (g_l - 1) g_l\xi-3 g_l^3 - 11 g_l^2  + 18 \right) 
\\  
&&
+ e g_l g_m \left(3 g_l^2 - 4 g_l + 9 \right)  
+ (3 g_l + 1) g_l^2 g_m^2\Big]; \nonumber \\ 
\Gamma^{(4)}_4 &=& -2 i e^2 \delta_4 + \frac{ie^2}{48 \pi^2 \epsilon}\Big[  e^2 \left(3 g_l - 6 \xi + 14 \right)+ e \left( g_m - 3 g_l g_m \right) 
+ (2 - 3 g_l) g_m^2 \Big].
\end{eqnarray}
\end{subequations}

The counterterms  $\delta_4$ e $\delta_6$ are determined by imposing the finiteness condition on the equation above. The counterterms associated with the four-point functions can be explicitly written as
\begin{subequations}
    \begin{eqnarray}
   \delta_4 &=& \frac{1}{96 \pi^{2} \epsilon}\Big[e^2\!\left(3 g_l - 6 \xi + 14\right) + e\left(g_m - 3 g_l g_m\right) + (2 - 3 g_l) g_m^{2}\Big] ;\\
   \delta_6 &=& \frac{1}{32 \pi^2 g_l \epsilon}\Big[e^2 \left(g_l^2 - 2 g_l \xi + 6\right) - e (g_l - 3) g_l g_m + g_l^2 g_m^2\Big];\\
   \delta_{\lambda} &=& \frac{1}{96 \pi^2 \epsilon}\Big[ 22 e^4 - 56 e^3 g_m - 6 e^2 \left(2 (g_l - 6) g_m^2 + 5 \lambda + 2 \lambda \xi \right) + e \big(8 (6 g_l + 5) g_m^3 \nonumber \\ \nonumber 
&&- 30 (g_l + 1) g_m \lambda \big) + 8 g_l^2 g_m^4 - 2 \lambda \big(3 (2 g_l^2 + g_l + 2) g_m^2 + (2 g_l^2 - g_l + 5) \tilde{\lambda} \big)+ 4 g_l^2 g_m^2 \tilde{\lambda} \\  
&& + (5 g_l^2 + 2 g_l + 17) \lambda^2 + g_l^2 \tilde{\lambda}^2 
+ 8 g_l g_m^4 - 8 g_l g_m^2 \tilde{\lambda} + 4 g_l \tilde{\lambda}^2 
+ 6 g_m^4 + 4 g_m^2 \tilde{\lambda} + 7 \tilde{\lambda}^2 \Big];\\
    \delta_{\tilde{\lambda}} &=& -\frac{1}{384 \pi^2 \epsilon}\Big[ 32 e^4 + 128 e^3 g_m + 24 e^2 \big( g_m^2(2 g_l + 6 ) + 2 \lambda -  \tilde{\lambda}(3- 2\xi)    \big)\nonumber \\  
&& + 8 e g_m \big(4 (3 g_l + 1) g_m^2 + 6 (g_l + 1) \lambda - 9 (g_l + 1) \tilde{\lambda} \big) + 4 (7 g_l^2 - 2 g_l + 3) g_m^4 \nonumber\\
&&+ (g_l - 1)^2 \lambda^2
- 8 (5 g_l^2 - g_l + 5) g_m^2 \tilde{\lambda} - 4 (2 g_l^2 + 5 g_l + 11) \lambda \tilde{\lambda}
\nonumber\\ 
&&
+ 2 (7 g_l^2 - 2 g_l + 19) \tilde{\lambda}^2 + 12 (g_l + 1)^2 g_m^2 \lambda\Big];
\end{eqnarray}
\end{subequations}

With all counterterms evaluated, we can proceed to determine the renormalization group functions. The beta functions for the coupling constants of the model are calculated using the counterterms along with the following relationships between the bare and renormalized couplings:
\begin{subequations}\label{Renormalizations}
\begin{eqnarray}\label{eq:renormalization}
e_0   &=& e\mu^{\epsilon} (1+\delta_3)^{-1/2};\\ \label{eq:bareadimensional}
\frac{Z_2}{{g_l}_0}&=& \frac{(1+\delta_{g_l})}{g_l};\\ \label{eq:renormalambda}
\lambda_0 Z_2^2 &=& \mu^{2\epsilon} (\lambda + \delta_\lambda);\\ \tilde{\lambda}_0 Z_2^2 &=& \mu^{2\epsilon}(\tilde{\lambda} + \delta_{\tilde{\lambda}})\\
g_{m0} Z_3^{1/2} Z_2 &=& \mu^{\epsilon}(g_m + \delta_{g_m})
\end{eqnarray}
\end{subequations}
where, in the first relation, we used $Z_{1}=Z_{2}$, as follows from Eqs.~\eqref{eq:delta2} and \eqref{eq:delta1}.

In contrast to the couplings $e$, $g_m$, $\lambda$, and $\tilde{\lambda}$, the bare definition of $g_l$ in Eq.~\eqref{eq:bareadimensional} does not require a compensating factor of $\mu^{\epsilon}$. This traces back to the canonical dimensionality of the operator it multiplies, $\frac{1}{2g_l}\,(\partial^\mu B_\mu)^2$. Indeed, if $B_\mu$ carries the standard Maxwell-type kinetic term, then in $D$ spacetime dimensions one has $[B_\mu]=D/2-1$, so that $[\partial\!\cdot\!B]=D/2$ and consequently $[(\partial\!\cdot\!B)^2]=D$. Hence the coefficient $1/g_l$ is canonically dimensionless for arbitrary $D$, and $g_l$ remains a marginal coupling under analytic continuation within dimensional regularization. Accordingly, the scale dependence of $g_l$ is entirely induced by UV renormalization, i.e., by its $\beta$ function, rather than by an explicit $\mu$-prefactor. In comparison, couplings that are marginal only in four dimensions --- such as the gauge coupling $e$ (and likewise $g_m$ in the present model) --- acquire a nonvanishing canonical mass dimension upon continuation to $D=4-2\epsilon$, and therefore must be accompanied by an explicit factor of $\mu^\epsilon$ so that the action remains dimensionless in dimensional regularization.

The relationships \eqref{Renormalizations}, together with the expressions for the counterterms, lead to the following beta functions:
\begin{subequations}\label{Eq:betas}
    \begin{eqnarray}
\beta(e) &=& \frac{e \left(e^2 (3 g_l - 5) - 12 e g_m - 3 (g_l + 1)  g_m^2\right)}{96 \pi^2};\\
\beta(\lambda) &=& \frac{1}{48 \pi^2} \Big[22 e^4- 56 e^3 g_m- 2 e^2 \big( 6 (g_l - 6) g_m^2 + (3 g_l + 29) \lambda \big)+ 4 (g_l - 1)^2 g_m^2 \tilde{\lambda}\nonumber \\ \nonumber
&&+ 8 e g_m \big( (6 g_l + 5) g_m^2 \nonumber - (3 g_l + 4) \lambda \big)- 4 (3 g_l^2 + 4) g_m^2 \lambda- 2 (2 g_l^2 - g_l + 5) \lambda \,\tilde{\lambda}\\ 
&&+ \big( 8 g_l (g_l + 1) + 6 \big) g_m^4+ \big( g_l (5 g_l + 2) + 17 \big) \lambda^2+ \big( g_l (g_l + 4) + 7 \big) \tilde{\lambda}^2\Big];\\ 
\beta(\tilde{\lambda}) &=&-\frac{1}{192 \pi ^2}\Big[
32 e^4
+128 e^3 g_m
+8 e^2 \big(6 (g_l+3) g_m^2+(3 g_l+5) \tilde{\lambda}+6 \lambda \big)\nonumber  \\ 
&&+16 e g_m \big((6 g_l+2) g_m^2+3 (g_l+1) \lambda -2 (3 g_l+2) \tilde{\lambda}\big)
+4 \left(7 g_l^2-2 g_l+3\right) g_m^4 \nonumber\\ \nonumber
&&-4 \lambda  \big((2 g_l^2+5 g_l+11) \tilde{\lambda}-3 (g_l+1)^2 g_m^2\big)
-8 \left(5 g_l^2+2 g_l+3\right) g_m^2 \tilde{\lambda}\\ 
&&+2 \big(7 g_l^2-2 g_l+19\big) \tilde{\lambda}^2
+(g_l-1)^2 \lambda ^2
\Big]
;\\
\beta(g_m) &=& \frac{1}{192 \pi^2}\Big[-20 e^3-12 e^2 (g_l + 2)\, g_m- 2 e (20 g_m^2 - 3\lambda - 6\tilde{\lambda})
\nonumber \\ 
&&
- 2 (3 g_l + 5) g_m^3 
+ 3 (g_l + 1) g_m \lambda+ 6 (g_l + 1) g_m \tilde{\lambda}\Big];\\
\beta(g_{l}) &=& \frac{1}{24 \pi^2}\left[e^2 (7 g_l-9 ) - 4 e g_l g_m +  (1-3 g_l)g_l g_m^2\right].
\end{eqnarray}
\end{subequations}

At this stage, the UV structure of the complex bumblebee model is fully determined at one loop within the MS scheme. Collecting the counterterms extracted from the two-, three-, and four-point functions, we obtained the complete set of renormalization constants required to renormalize the gauge, longitudinal, magnetic-type, and quartic sectors, and we derived the corresponding renormalization-group functions in Eqs.~(\ref{Eq:betas}a)--(\ref{Eq:betas}e). As expected, the resulting beta functions are independent of the gauge-fixing parameter, whereas gauge-parameter dependence may enter only through anomalous dimensions. 

It is worth stressing that the hypersurface in coupling space defined by $g_m=g_l=\lambda=\tilde\lambda=0$ is not stable under renormalization once the gauge interaction is switched on. Indeed, Eqs.~(\ref{Eq:betas}b)-(\ref{Eq:betas}d) exhibit genuine ``source'' terms driven solely by the electric charge: $\beta(\lambda)$ contains an additive contribution $\propto e^{4}$, $\beta(\tilde\lambda)$ contains an additive contribution $\propto e^{4}$, and $\beta(g_m)$ contains an additive contribution $\propto e^{3}$ even at $g_m=g_l=\lambda=\tilde\lambda=0$. Concretely, setting $g_m=g_l=\lambda=\tilde\lambda=0$ in Eqs.~(\ref{Eq:betas}b)-(\ref{Eq:betas}d) yields
\begin{equation}\label{eq:betas_sources}
\beta(\lambda)=\frac{11\,e^{4}}{24\pi^{2}},\qquad
\beta(\tilde\lambda)=-\frac{\,e^{4}}{6\pi^{2}},\qquad
\beta(g_m)=-\frac{5\,e^{3}}{48\pi^{2}},\qquad \beta(g_l)=-\frac{3e^2}{8\pi^2},
\end{equation}
so that gauge fluctuations radiatively regenerate the quartic self-couplings and the magnetic-type interaction even if they are tuned to vanish at some reference scale. From an effective-field-theory viewpoint this reflects operator mixing among the marginal interactions compatible with Lorentz symmetry and the $U(1)$ phase symmetry of $B_\mu$: the gauge sector acts as a driver for a nontrivial RG flow in the full coupling space, and a consistent renormalization therefore requires including $(g_l,g_m,\lambda,\tilde\lambda)$ as running couplings rather than treating them as optional deformations.

These RG functions, Eqs. \eqref{Eq:betas}, provide the essential input for the resummation of leading logarithms and for a controlled discussion of radiative symmetry breaking. In the next section, we therefore turn to the Vilkovisky--DeWitt formulation of the effective potential and develop an RG-improvement scheme in RG-covariant (normal) field coordinates, where the RG equation is defined solely by the beta functions.

\section{Renormalization group improvement for the Vilkovisky-DeWitt Effective Potential}

The study of the effective action provides a powerful framework for encoding the quantum behavior of physical systems in terms of classical fields. In particular, the associated effective potential offers a natural setting convenient to analyze spontaneous symmetry breaking and the dynamical generation of mass. In gauge theories, however, the gauge dependence of the conventional effective potential obscures the physical interpretation of radiative symmetry breaking and vacuum stability analyses, especially in the CW framework.

Concretely, the standard one-particle-irreducible (1PI) effective action can be written schematically as
\begin{eqnarray}\label{seff}
\Gamma[\phi]
&=& \int d^4x\,\mathcal{L}[\phi]
+ \frac{i}{2}\ln\det\!\left[ -\frac{\delta^2\mathcal{L}}{\delta\phi\,\delta\phi}\right]
+ \cdots
= \int d^4x\, V_{\text{eff}}(\phi),
\end{eqnarray}
where the ellipsis $(\cdots)$ denotes higher-order quantum contributions. When obtained from the standard gauge-fixed path integral, $\Gamma[\phi]$ depends explicitly on the gauge-fixing condition and on the particular choice of field coordinates in field space~\cite{Jackiw:1974cv}. Consequently, the ordinary effective potential $V_{\text{eff}}(\phi)$ is gauge dependent away from its stationary points; gauge invariance is recovered only at its extrema, as encoded in the Nielsen identity~\cite{Nielsen:1975fs} (for explicit verifications in scalar electrodynamics and for discussions of gauge invariance in the presence of daisy resummations in CW-type models, see Refs.~\cite{Bazeia:1988pz,deLima:1989yf}). In general, the RG equation obeyed by $V_{\text{eff}}$ involves the anomalous dimension of the field, which in gauge theories is typically gauge dependent itself,
\begin{equation}\label{rge_veff}
\left(
\mu\frac{\partial}{\partial \mu}
+ \beta_i\frac{\partial}{\partial \lambda_i}
- \gamma_\phi\,\phi\,\frac{\partial}{\partial \phi}
\right)
V_{\text{eff}}(\phi,\lambda_i,\mu) = 0.
\end{equation}

As shown by Vilkovisky~\cite{Vilkovisky:1984st}, this gauge dependence arises because the second functional derivative $S_{,AB}$ of the classical action $S[\phi]$ mixes physical and gauge directions in field space. The Vilkovisky--DeWitt (VDW) construction removes these ambiguities by endowing the space of fields $\{\phi^A(x)\}$ with a covariant geometric structure and by defining the effective action as an unambiguous scalar on this manifold.

In order to characterize the geometry of field space, let us define $G_{AB}(\phi)$ as a (DeWitt) metric on the field space that defines the inner product of variations $\delta\phi^A$, and let $K^A{}_\alpha(\phi)$ denote the generators of gauge transformations, so that an infinitesimal transformation acts as $\delta_\epsilon\phi^A = K^A{}_\alpha\,\epsilon^\alpha$. The induced metric on the space of gauge orbits (the dependence of the effective action on this metric has earlier been studied in \cite{Odintsov:1991yx}) is then
\begin{equation}
\omega_{\alpha\beta}(\phi)
= K^A{}_\alpha(\phi)\,G_{AB}(\phi)\,K^B{}_\beta(\phi),
\end{equation}
and the projector onto directions \emph{horizontal} to the gauge orbits (i.e., orthogonal to gauge directions) is given by
\begin{equation}
\Pi^A{}_B
= \delta^A{}_B
- K^A{}_\alpha\,(\omega^{-1})^{\alpha\beta}\,K_{B\beta},
\qquad
K_{B\beta} \equiv G_{BC}K^C{}_\beta.
\end{equation}
The VDW connection is chosen so that covariant derivatives do not mix physical and gauge directions,
\begin{equation}
\bar\Gamma^{A}{}_{BC}
= \Gamma^{A}{}_{BC}[G]
+ K^A{}_\alpha\,(\omega^{-1})^{\alpha\beta}\,\nabla_{(B}K_{C)\beta},
\end{equation}
where $\Gamma[G]$ is the Levi--Civita connection associated with $G_{AB}$ and $\nabla$ denotes the corresponding covariant derivative.

The VDW effective action is defined by replacing the ordinary second functional variations by the \emph{covariant} Hessian $S_{;AB}\equiv\nabla_A\nabla_B S$ and restricting the functional traces to the horizontal subspace:
\begin{equation}
\Gamma_{_{\rm VDW}}[\phi]
= S[\phi]
+ \frac{i}{2}\,\mathrm{Tr}_{\mathcal H}\!\ln\!\big(S_{;AB}\big)
- \frac{i}{2}\,\mathrm{Tr}\ln(\omega_{\alpha\beta})
+ \cdots,
\label{eq:GammaVDW-oneloop}
\end{equation}
where $\mathrm{Tr}_{\mathcal H}$ indicates that the trace is taken with the projector $\Pi$ inserted in the propagators, and the last term accounts for the gauge-orbit measure. Equation~\eqref{eq:GammaVDW-oneloop} represents the one-loop approximation to the unique effective action in the sense of Vilkovisky and DeWitt; higher-loop contributions are obtained by the usual loop expansion with VDW-covariant vertices and propagators. By construction, $\Gamma_{_{\rm VDW}}$ is independent of the gauge-fixing condition and of field reparametrizations \emph{off shell} (for a review of this effective action, see \cite{Odintsov:1989gz,Buchbinder:1992gdx}).

For constant backgrounds (spacetime-independent fields), the effective action reduces to
\begin{equation}
\Gamma_{_{\rm VDW}}[\phi_{\rm c}]
= \int d^D x\,V_{_{\rm VDW}}(\phi_{\rm c}),
\end{equation}
which defines the VDW effective potential $V_{_{\rm VDW}}$. Operationally, $V_{_{\rm VDW}}$ is obtained by evaluating the functional traces in Eq.~\eqref{eq:GammaVDW-oneloop} at constant fields, with the covariant Hessian projected onto the physical (horizontal) subspace. At the one-loop order this yields the familiar sum over \emph{physical} modes only,
\begin{equation}
V_{_{\rm VDW}}^{(1)}(\phi_{\rm c})
= \frac{i}{2}\int\!\frac{d^D p}{(2\pi)^D}\,
\sum_{n\in{\rm phys}}\ln\!\big(p^2 + m_n^2(\phi_{\rm c}) - i0\big),
\end{equation}
where $\{m_n^2(\phi_{\rm c})\}$ are the field-dependent eigenvalues of $S_{;AB}$ restricted by $\Pi$. Unphysical Goldstone, longitudinal, and ghost contributions cancel identically once the VDW geometry is implemented. Consequently, $V_{_{\rm VDW}}$ is gauge- and parametrization-independent for all field values (and not only at its extrema); for explicit examples, see Ref.~\cite{Kunstatter:1986qa,Lavrov:1988is,Odintsov:1989sx}.

While the VDW construction provides a unique gauge- and parametrization-independent effective action, a fully systematic RG-improved treatment of the corresponding effective potential, in the spirit of CW, has not been developed. In the following sections we will develop the RGE improvement method to compute the VDW effective potential and apply it in the obtaining the VDW potential for the complex bumblebee field.

\subsection{Renormalization group equation for VDW effective potential}\label{sec:RGcov}

In the VDW formulation the effective action $\Gamma_{_{\rm VDW}}$ is a scalar on field space. Its covariant RG equation can be written as
\begin{equation}
\label{eq:RGE-cov-phi}
\left(
\mu\frac{\partial}{\partial\mu}
+\sum_i \beta_i(\lambda)\frac{\partial}{\partial\lambda_i}
-\!\!\int d^Dx~\gamma^{\ A}{}_{B}(\lambda)\,\phi^B(x)\,\nabla^{(\phi)}_{A}
\right)\Gamma_{_{\rm VDW}}[\phi;\lambda,\mu]=0,
\end{equation}
where $\{\lambda_i\}$ denotes the set of renormalized couplings and masses,
$\gamma^{\ A}{}_{B}$ is the anomalous--dimension tensor, and
$\nabla^{(\phi)}_{A}$ is the VDW covariant functional derivative in field space. The last term in~\eqref{eq:RGE-cov-phi} is the familiar field-rescaling piece. We now show that in \emph{normal geodesic coordinates} -- the RG--covariant field variables $\chi$ -- this term is absorbed into the definition of the $\mu$-derivative at fixed $\chi$.

Let $\bar\phi$ be a background configuration and define the normal geodesic
coordinates $\chi^A$ via the exponential map on field space,
\begin{equation}
\label{eq:expmap}
\phi \;=\; {\rm Exp}_{\bar\phi}(\chi), 
\qquad
\bar\Gamma^{A}{}_{BC}(\bar\phi)=0,
\qquad
\nabla^{(\phi)}_{A}\xrightarrow[\phi\to\bar\phi]{}\frac{\partial}{\partial\chi^A}.
\end{equation}
By construction, $\chi$ transforms homogeneously under field reparametrizations and gauge, and acts as the RG--covariant field variable in a neighborhood of $\bar\phi$.

We define the field RG flow at fixed $\chi$ by
\begin{equation}
\label{eq:RGflow-phi}
\Big(\mu\frac{d\phi^A}{d\mu}\Big)_{\!\chi} \;=\; -\,\gamma^{\ A}{}_{B}\,\phi^B.
\end{equation}
Equation~\eqref{eq:RGflow-phi} states that if the RG--covariant coordinate
$\chi$ is held fixed, then the renormalized field $\phi$ runs exactly according to its anomalous dimension.

For any functional $F[\phi,\lambda,\mu]$ we define the total derivative at
fixed $\chi$,
\begin{equation}
\label{eq:Dmuchi-def}
\Big(\mu\tfrac{d}{d\mu}\Big)_{\!\chi} F
\;\equiv\;
\mu\partial_\mu F
+\sum_i \beta_i\,\partial_{\lambda_i} F
+\!\int d^Dx\,
\frac{\delta F}{\delta\phi^A(x)}
\Big(\mu\tfrac{d\phi^A(x)}{d\mu}\Big)_{\!\chi}.
\end{equation}
Using~\eqref{eq:RGflow-phi} and replacing ordinary by covariant derivatives in field space, we obtain the \emph{operator identity}
\begin{equation}
\label{eq:op-identity}
\Big(\mu\tfrac{d}{d\mu}\Big)_{\!\chi}
\;=\;
\mu\partial_\mu
+\sum_i \beta_i\,\partial_{\lambda_i}
-\!\!\int d^Dx~\gamma^{\ A}{}_{B}\,\phi^B\,\nabla^{(\phi)}_{A}.
\end{equation}

Applying the identity~\eqref{eq:op-identity} to the covariant RG equation~\eqref{eq:RGE-cov-phi} yields
\begin{equation}
\label{eq:RGE-chi-master}
\Big(\mu\tfrac{d}{d\mu}\Big)_{\!\chi}\,
\Gamma_{_{\rm VDW}}[\phi(\chi);\lambda,\mu]\;=\;0.
\end{equation}
In normal coordinates at the background, where
$\nabla^{(\phi)}\!\to\!\partial/\partial\chi$, Eq.~\eqref{eq:RGE-chi-master}
reduces to
\begin{equation}
\label{eq:RGE-chi-compact}
\left(
\mu\frac{\partial}{\partial\mu}
+\sum_i \beta_i(\lambda)\frac{\partial}{\partial\lambda_i}
\right)\Gamma_{_{\rm VDW}}[\chi;\lambda,\mu]=0,
\end{equation}
i.e. the anomalous- dimension term does not appear explicitly when the RG
operator is taken at fixed RG--covariant field $\chi$ (a more detailed derivation is found in the Appendix). 

For constant backgrounds and the same algebra gives
\begin{equation}
\label{eq:RGE-V-chi}
\left(\mu\frac{\partial}{\partial\mu}
+\sum_i\beta_i(\lambda)\frac{\partial}{\partial\lambda_i}\right)
V_{_{\rm VDW}}(\chi,\lambda,\mu)=0,
\end{equation}
i.e. the RG equation for the unique (gauge/parametrization--independent)
VDW potential in normal geodesic coordinates, with no explicit anomalous dimension term.

In a conventional 1PI formulation one may have a term $\beta_\xi\,\partial_\xi$ in the RG operator. The Nielsen identity converts $\partial_\xi$ into a field direction and it combines with the anomalous--dimension piece exactly as above, so that in RG--covariant variables the reduced operator is again $\mu\partial_\mu+\sum\limits_i\beta_i\partial_{\lambda_i}$. In the VDW formalism,
$\partial_\xi\Gamma_{_{\rm VDW}}=0$ off-shell, and no $\beta_\xi$ term appears.

\subsection{Leading logarithms for the VDW potential}\label{LL-RGE}

In classically scale-invariant theories, it is important to investigate whether spontaneous symmetry breaking can occur dynamically via the CW mechanism~\cite{Coleman:1973jx}. The effective potential is one of the primary tools for addressing this question. Among the available techniques, the computation of the LL effective potential using RG improvement is particularly powerful~\cite{McKeon:1998tr}. This procedure provides a systematic framework for resummation of logarithmic corrections and obtaining the LL approximation to the potential. The approach has proven remarkably effective in a variety of field-theoretical contexts, as demonstrated in Refs.~\cite{Elias:2003zm,Elias:2004bc,Dias:2010it,Lehum:2019msl,Souza:2020hjd,Lehum:2023qnu}.

However, the usual treatment based on Eq.~\eqref{rge_veff} can be gauge dependent, since the anomalous dimension is in general gauge dependent in gauge theories. Instead, we will develop an RG improvement for the VDW potential, which is independent of gauge fixing and of field reparametrizations \emph{off shell}.

In four dimensions, the marginal operator in the tree-level potential is proportional to $\chi^4$,
\begin{eqnarray}\label{eqV0vdw}
{V_0}_{_{\rm VDW}} &=& \frac{\lambda}{4!}\,\chi^4,
\end{eqnarray}
where $\chi$ is the RG-invariant field. On dimensional grounds, the perturbative expansion of the VDW potential takes the general form
\begin{eqnarray}
V_{_{\rm VDW}}(\chi)
= A_0(x)\,\chi^4 + A_1(x)\,\chi^4 L + A_2(x)\,\chi^4 L^2 + \cdots,
\end{eqnarray}
where $x$ collectively denotes the coupling constants, $L \equiv \ln(\chi^2/\mu^2)$, and $\mu$ is the renormalization scale introduced through dimensional regularization. Each coefficient
$A_i(x)=a_1^{(i)}x+a_2^{(i)}x^2+a_3^{(i)}x^3+\cdots$ represents a power series in the couplings, whose order corresponds to the number of loops in the perturbative expansion.

It is often convenient to reorganize the potential into a series of  logarithmic contributions:
\begin{eqnarray}\label{eq:ansatz}
V_{_{\rm VDW}}(\chi)
= \chi^4\!\left[
\sum_{n=0}^{\infty} C_{LL}^{(n)}(x)\,L^n
+ \sum_{n=0}^{\infty} C_{NLL}^{(n)}(x)\,L^{n}
+ \cdots
+ \frac{\delta_\lambda}{4!}
\right],
\end{eqnarray}
where $C_{LL}^{(n)}(x)=a_{n+1}^{(n)}x^{n+1}$ and $C_{NLL}^{(n)}(x)=a_{n+2}\,x^{n+2}$ represent the leading- and next-to-leading-logarithmic coefficients, respectively. The quantity $\delta_\lambda$ is the counterterm fixed by the CW renormalization condition,
\begin{eqnarray}\label{cw:ren_cond}
\frac{\partial^4 V_{_{\rm VDW}}}{\partial \chi^4}\Bigg|_{\chi=\mu}
= \frac{\partial^4 V_{0_{_{\rm VDW}}}}{\partial \chi^4}\Bigg|_{\chi=\mu},
\end{eqnarray}
where we set $\mu$ as the renormalization scale.

The VDW potential must satisfy the RG equation~\eqref{eq:RGE-V-chi},
\begin{eqnarray}\label{eq:RGE}
\left(
\mu\frac{\partial}{\partial\mu}
+ \beta_x\frac{\partial}{\partial x}
\right)V_{_{\rm VDW}} = 0,
\end{eqnarray}
where $\beta_x \equiv \mu\,\mathrm{d}x/\mathrm{d}\mu$ denotes the beta function of the (collective) coupling $x$. Using the identity
$\mu\,\partial V_{_{\rm VDW}}/\partial\mu = -2\,\partial V_{_{\rm VDW}}/\partial L$, one obtains the compact form
\begin{eqnarray}\label{eq:RGEcompact}
\left[
-2\frac{\partial}{\partial L}
+ \beta_x\frac{\partial}{\partial x}
\right]V_{_{\rm VDW}} = 0.
\end{eqnarray}

Substituting the ansatz~\eqref{eq:ansatz} into Eq.~\eqref{eq:RGEcompact} and isolating the LL contributions yields the recursion relation
\begin{eqnarray}\label{recursive_relation}
C_{LL}^{(n)}(x)
= \frac{1}{2n}
\left(\beta_x\frac{\partial}{\partial x}\right) C_{LL}^{(n-1)}(x),
\qquad n \ge 1.
\end{eqnarray}

At the one-loop level, only the first correction term $C_{LL}^{(1)}(x)$ is required. To reproduce the tree-level potential, we set $C_{LL}^{(0)}(x) = V_{0_{_{\rm VDW}}}/\chi^4$, and $C_{LL}^{(1)}(x)$ is then obtained from Eq.~\eqref{recursive_relation}. The resulting VDW potential at one loop reads
\begin{eqnarray}
V_{_{\rm VDW}}
= V_{0_{_{\rm VDW}}}
+ \frac{\delta_\lambda}{4!}\,\chi^4
+ \frac{1}{2} C^{(1)}_{LL}(x)\,\chi^4
\ln\!\left(\frac{\chi^2}{\mu^2}\right),
\end{eqnarray}
where $\delta_\lambda$ is a finite counterterm determined by the CW renormalization condition~\eqref{cw:ren_cond}.

The conditions for the minimum of the VDW potential are
\begin{subequations}
\begin{eqnarray}
\frac{\partial V_{_{\rm VDW}}}{\partial \chi}\Bigg|_{\chi=\mu} &=& 0, \label{cond_gap}
\\[4pt]
\frac{\partial^2 V_{_{\rm VDW}}}{\partial \chi^2}\Bigg|_{\chi=\mu} &=& m_{\chi}^2 > 0, \label{cond_mass}
\end{eqnarray}
\end{subequations}
where the first condition fixes the renormalization scale $\mu$ to coincide with the minimum of the VDW potential, and the second one determines the radiatively induced mass for the $\chi$ field.

It is important to emphasize the domain of validity of the RG-improved VDW potential constructed here. Our analysis is perturbative and relies on one-loop beta functions together with a LL resummation, so the resulting $V_{_{\rm VDW}}^{LL}(\chi)$ is reliable only as long as the couplings remain in the weak-coupling regime and the logarithms are not large enough to invalidate the truncation at LL order. In practice, the approximation is most trustworthy for field values $\chi$ of order the renormalization scale $\mu$, where $L=\ln(\chi^2/\mu^2)$ is moderate, and well below any Landau-pole scale of the theory. In addition, in a generic curved field manifold, Riemann normal coordinates provide only a locally linear parametrization around a chosen reference configuration, with nontrivial curvature effects entering at higher orders in the geodesic distance. Finally, as in the conventional 1PI approach, the RG-improved potential should be regarded as an approximate description of the vacuum structure rather than as the exact convex effective potential: it accurately captures the radiatively induced minimum and the associated mass scale, but it is not expected to provide a quantitatively precise global description for arbitrarily large field values or in strongly coupled regimes.

In order to use Eqs.~\eqref{eq:ansatz} and~\eqref{recursive_relation} to compute the LL VDW potential, it is important to note that the RG functions differ slightly between the CW and $\overline{\mathrm{MS}}$ schemes~\cite{Chishtie:2007vd}. This distinction plays no role at one-loop order, and thus does not affect the LL VDW potential. However, it must be taken into account when computing the next-to-leading-logarithmic contributions, for which the two-loop beta functions are required.

\section{THE VDW EFFECTIVE POTENTIAL FOR THE BUMBLEBEE COMPLEX FIELD}

Having examined the UV behavior of the model, we now turn to investigate whether LSB can occur dynamically via the CW mechanism~\cite{Coleman:1973jx}. To this end,
we now apply it to a real constant background of the complex bumblebee model. Our purpose is to extract the corresponding RG-improved one-loop LL VDW potential and to analyze whether it supports a radiatively induced LSB vacuum.

It is important to emphasize that the classical potential must be expressed in terms of the real components of the complex bumblebee field, 
$B^\mu = B_1^\mu + i B_2^\mu$, with $(B^\mu)^{*} = B_1^\mu - i B_2^\mu$, where $B_1^\mu$ and $B_2^\mu$ are real vector fields.  
In this representation, the classical potential can be rewritten as
\begin{eqnarray}
V_0 &=& \frac{\lambda}{4}\,(B^{*\mu}B_{\mu})^2
      +\frac{\tilde{\lambda}}{4}\,(B^{*\mu}B^{\nu}B^*_{\mu}B_{\nu}) \nonumber\\
&=& \frac{\lambda+\tilde{\lambda}}{4}\left(B_1^4+B_2^4\right)
   +\frac{\lambda-\tilde{\lambda}}{2}\,B_1^2 B_2^2
   +\tilde{\lambda}\,(B_1\!\cdot\!B_2)^2,
\end{eqnarray}
where $B_i^2 \equiv B_i^\mu B_{i\mu}$ $(i=1,2)$.  
The effective potential is then evaluated for a real constant background configuration,  
$\langle B^\mu\rangle = \langle B_1^\mu + i B_2^\mu \rangle = B_c^\mu$.

In four dimensions, the marginal operator in the tree-level potential is proportional to $B^4$,
\begin{equation}
    V_{0_{\text{VDW}}}=\frac{\lambda+\tilde{\lambda}}{4}B_c^4\equiv\frac{\lambda_r}{4}B_c^4.
\end{equation}

On dimensional grounds, the perturbative expansion of the effective potential takes the general structure
\begin{eqnarray}
V_{\text{eff}}(B_c^2)=A_0(x)B_c^4 + A_1(x)B_c^4L + A_2(x)B_c^4L^2 + \cdots,
\end{eqnarray}
where $x$ collectively denotes the coupling constants, $L=\ln(B_c^2/\mu^2)$, $B_c^\mu$ is the classical background field, and $\mu$ is the renormalization scale introduced through dimensional regularization. Each coefficient $A_i(x)=a_0^{(i)}x+a_1^{(i)}x^2+a_2^{(i)}x^3+\cdots$ represents a power series in the couplings, whose order corresponds to the number of loops in the perturbative expansion.

It is often convenient to reorganize the potential into a series of logarithmic contributions:
\begin{eqnarray}\label{eq:ansatz1}
V_{\text{eff}}(B_c^2)=B_c^4\!\left[\sum_{n=0}^{\infty}C_{LL}^{(n)}(x)L^n+\sum_{n=0}^{\infty}C_{NLL}^{(n)}(x)L^{n+1}+\cdots + \frac{\delta}{4}\right] ,
\end{eqnarray}
with $C_{LL}^{(n)}(x)=a_n^{(n)}x^{n+1}$ and $C_{NLL}^{(n)}(x)=a_{n+1}^{(n)}x^{n+2}$ representing the leading- and next-to-leading-log coefficients, respectively. The quantity $\delta$ is the counterterm fixed by
the CW renormalization condition,
\begin{eqnarray}\label{eq:RC}
\frac{d^4V_0}{dB_c^\mu d{B_c}_\mu dB_c^\nu d{B_c}_\nu}=\frac{d^4V_{\text{eff}}}{dB_c^\mu d{B_c}_\mu dB_c^\nu d{B_c}_\nu}\Bigg|_{B_c^2=\mu^2}=6\lambda_r,
\end{eqnarray}
where we set $\mu$ as the renormalization scale.

A crucial consequence is that the RGE for the VDW potential reduces to a form controlled solely by the beta functions of the couplings as we can see in the Eq.\eqref{eq:RGE-V-chi}.
The logarithmic ansatz for the potential, Eq.~\eqref{eq:ansatz1}, leads to the following LL recursive relation,
\begin{equation}\label{recursive_relation1}
C_{LL}^{(n)}(\lambda,\tilde{\lambda})
= \frac{1}{2n}
\left(
\beta_{\lambda}\frac{\partial}{\partial \lambda}
+ \beta_{\tilde{\lambda}}\frac{\partial}{\partial \tilde{\lambda}}
\right)
C_{LL}^{(n-1)}(\lambda,\tilde{\lambda}),
\qquad n \geq 1,
\end{equation}
with $C_{LL}^{(0)}(\lambda_r)
=\lambda_r/4$. Although both the VDW effective potential and the LL coefficient are, in principle, functions of all the couplings, they effectively depend only on $\beta_{\lambda}$ and $\beta_{\tilde{\lambda}}$. This is because $C_{LL}^{(0)}$ is determined by the tree-level potential, and therefore depends explicitly on $\lambda$ and $\tilde{\lambda}$.

The resulting one-loop effective potential then reads
\begin{eqnarray}
V_{\text{eff}}=\frac{(\lambda_r+\delta)}{4}B_c^4+C_{LL}^{(1)}(\lambda,\lambda_r)B_c^4\ln\!\left(\frac{B_c^2}{\mu^2}\right).
\end{eqnarray}

In order to determine its minimum, $V_{\text{eff}}$ has to satisfy the following stationary conditions:
\begin{subequations}\label{eq:stationary}
\begin{eqnarray}
\frac{dV_{\text{eff}}}{dB_c^\mu}&=&0\quad\text{for some }B_c^2=\mu^2,\label{eq:gap}\\
m_B^2&=&\frac{d^2V_{\text{eff}}}{dB_c^\mu dB_{c\,\mu}}\Big|_{B_c^2=\mu^2}>0.
\end{eqnarray}
\end{subequations}

The CW renormalization condition \eqref{eq:RC} fixes the counterterm to be
\begin{eqnarray}
\delta &=& -\frac{25}{2304\pi^{2}}\Bigg[
56 e^{4}
-352 e^{3} g_{m}
+4\left(3+10 g_{l}+g_{l}^{2}\right) g_{m}^{4}
+\left(53+2 g_{l}+17 g_{l}^{2}\right)\lambda^{2}
\nonumber\\
&&
+\left(24-12 g_{l}+12 g_{l}^{2}\right)\lambda\,\lambda_{r}
+\left(-10+20 g_{l}-10 g_{l}^{2}\right)\lambda_{r}^{2}
\nonumber\\
&&
+16 e g_{m}\Big((8+6 g_{l}) g_{m}^{2}-15(1+g_{l})\lambda+2(2+3 g_{l})\lambda_{r}\Big)
\nonumber\\         
&&
-8 e^{2}\Big(6(-3+2 g_{l}) g_{m}^{2}+30\lambda+(5+3 g_{l})\lambda_{r}\Big)
\nonumber\\
&&
-4 g_{m}^{2}\Big((29+2 g_{l}+29 g_{l}^{2})\lambda-2(5-2 g_{l}+7 g_{l}^{2})\lambda_{r}\Big)
\Bigg].
\end{eqnarray}

Using the minimization condition, we find the value of the minimum, which can be written perturbatively as an expansion in powers of the coupling constants for $\lambda_r$ 
\begin{eqnarray}
\lambda_r &=& \frac{11}{1152\pi^{2}}\Bigg[
56 e^{4}
-352 e^{3} g_{m}
+4\left(3+10 g_{l}+g_{l}^{2}\right) g_{m}^{4}
-4\left(29+2 g_{l}+29 g_{l}^{2}\right) g_{m}^{2}\lambda
\nonumber\\
&&
+\left(53+2 g_{l}+17 g_{l}^{2}\right)\lambda^{2}
-48 e^{2}\Big((-3+2 g_{l}) g_{m}^{2}+5\lambda\Big)
\nonumber\\
&&+16 e\Big((8+6 g_{l}) g_{m}^{3}-15(1+g_{l}) g_{m}\lambda\Big)
\Bigg]+\cdots.
\end{eqnarray}

In this process, the dimensionless parameter $\lambda_r$ is replaced by the mass scale $\mu$ generated by the vacuum, in a mechanism known as dimensional transmutation \cite{Coleman:1973jx}. Since $\lambda_r \sim \mathcal{O}(e^4,g_m^4,\cdots)$, terms of order $\lambda^2_r$ can be consistently neglected. 
 Accordingly, the one-loop VDW effective potential can be expanded up to $\mathcal{O}(e^4,\lambda^2,g_m^4)$ as
\begin{eqnarray}\label{eq:VCWoneloop}
  V_{\text{eff}} &=& \Bigg[
56 e^{4}
-352 e^{3} g_{m}
+4\left(3+10 g_{l}+g_{l}^{2}\right) g_{m}^{4}
-4\left(29+2 g_{l}+29 g_{l}^{2}\right) g_{m}^{2}\lambda
\nonumber\\
&&
+\left(53+2 g_{l}+17 g_{l}^{2}\right)\lambda^{2}
-48 e^{2}\Big((-3+2 g_{l}) g_{m}^{2}+5\lambda\Big)
\nonumber\\
&&+16 e\Big((8+6 g_{l}) g_{m}^{3}-15(1+g_{l}) g_{m}\lambda\Big)
\Bigg]\frac{B^{4}}{1536\pi^{2}}\Bigg(-\frac12+\ln\frac{B^{2}}{\mu^{2}}\Bigg).
\end{eqnarray}
The expression \eqref{eq:VCWoneloop} for the effective potential is valid only in the region of classical field space where $\ln(B_c^2/\mu^2)$ remains small. In particular, it is not reliable for very small or very large values of $B_c^2$.

\begin{figure}[ht!]
	\includegraphics[angle=0,width=15cm]{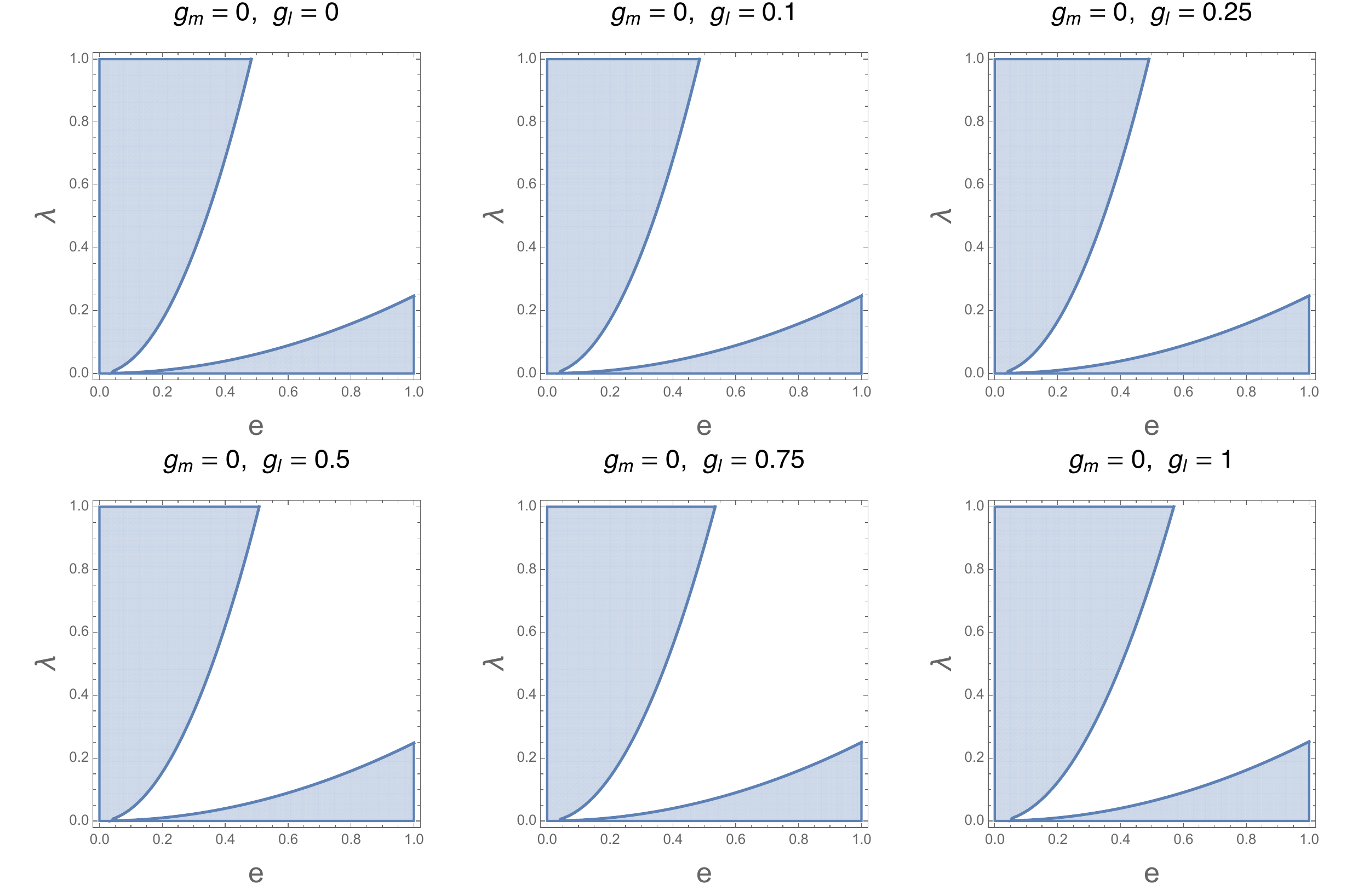}
	\caption{Two-dimensional slices of the parameter space $(e,\lambda,g_m,g_l)$ illustrating the domain where $m_B^2/\mu^2>0$. Each panel shows the $(e,\lambda)$ region (shaded) satisfying $m_B^2(e,\lambda;g_m,g_l)>0$ for a fixed choice of $(g_m,g_l)$ displayed in the panel label, with $0<e<1$ and $0<\lambda<1$. The thick contour denotes the locus $m_B^2=0$.}
	\label{ps01}
\end{figure}

\begin{figure}[ht!]
	\includegraphics[angle=0,width=15cm]{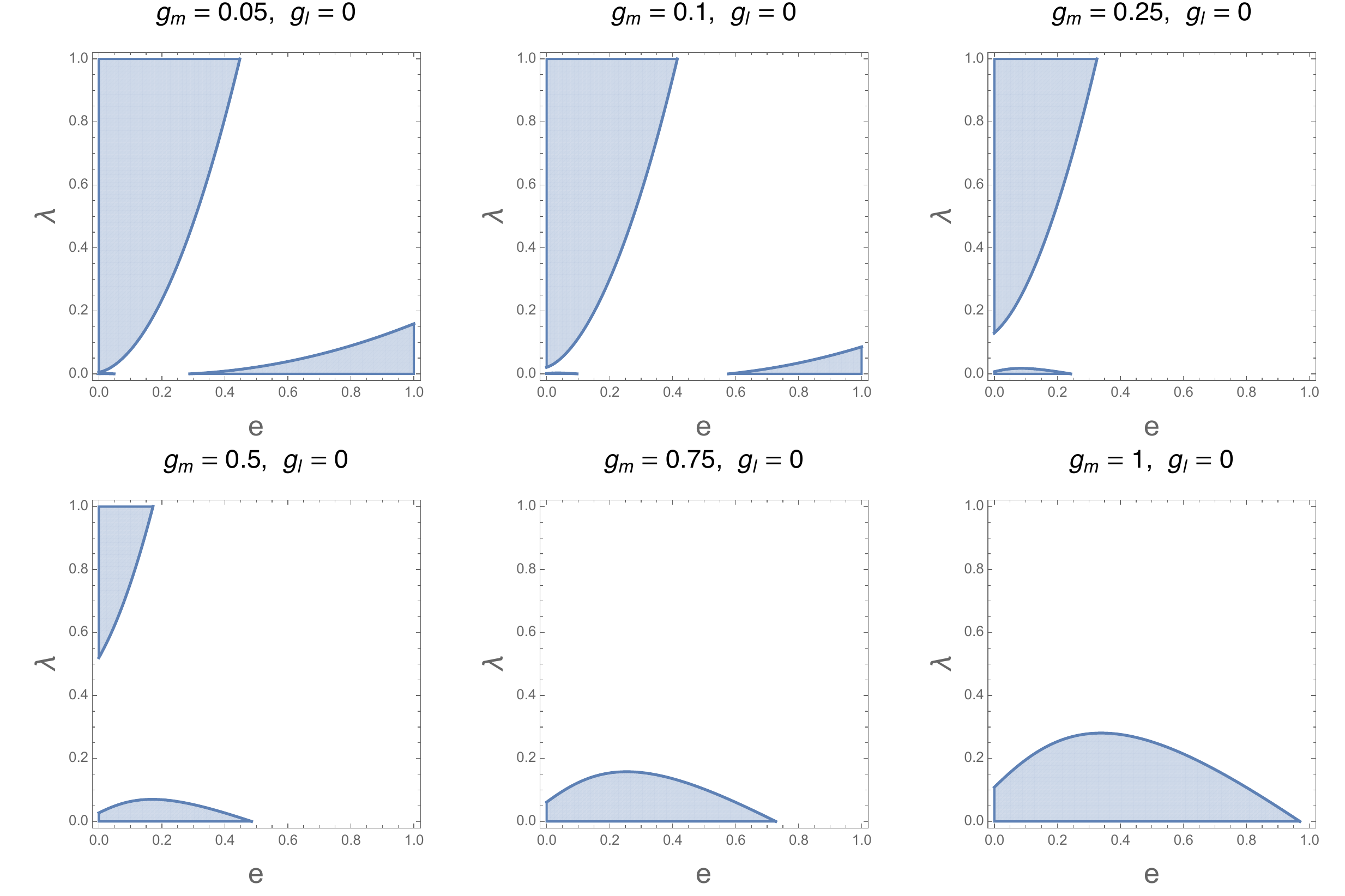}
	\caption{Two-dimensional slices of the parameter space $(e,\lambda,g_m,g_l)$ illustrating the domain where $m_B^2/\mu^2>0$. Each panel shows the $(e,\lambda)$ region (shaded) satisfying $m_B^2(e,\lambda;g_m,g_l)>0$ for a fixed choice of $(g_m,g_l)$ displayed in the panel label, with $0<e<1$ and $0<\lambda<1$. The thick contour denotes the locus $m_B^2=0$.}
	\label{ps02}
\end{figure}

\begin{figure}[ht!]
	\includegraphics[angle=0,width=15cm]{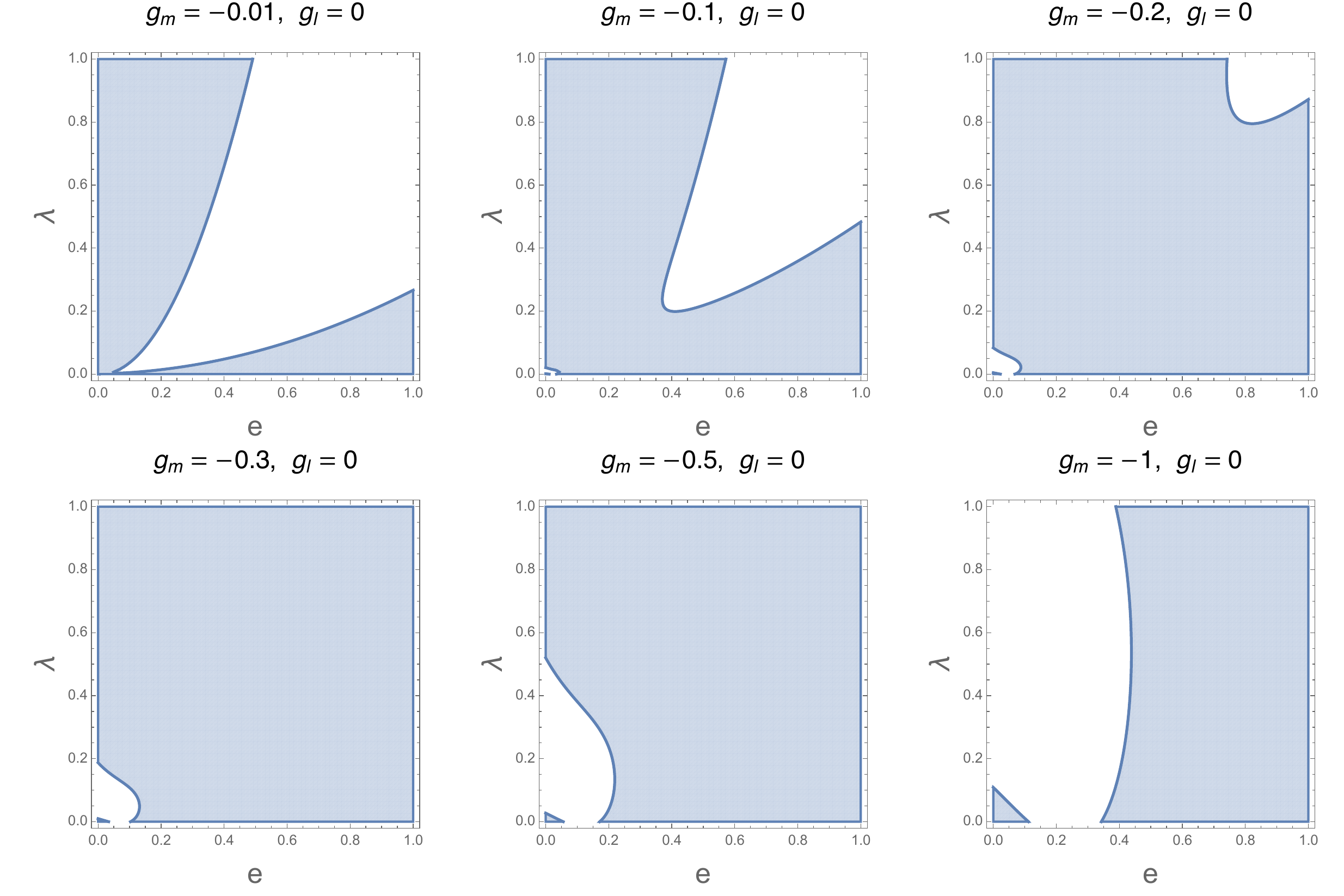}
	\caption{Two-dimensional slices of the parameter space $(e,\lambda,g_m,g_l)$ illustrating the domain where $m_B^2/\mu^2>0$. Each panel shows the $(e,\lambda)$ region (shaded) satisfying $m_B^2(e,\lambda;g_m,g_l)>0$ for a fixed choice of $(g_m,g_l)$ displayed in the panel label, with $0<e<1$ and $0<\lambda<1$. The thick contour denotes the locus $m_B^2=0$.}
	\label{ps03}
\end{figure}

From the stationarity conditions, Eqs.~(\ref{eq:stationary}a) and (\ref{eq:stationary}b), the bumblebee field acquires a dynamically generated mass. Explicitly,
 \begin{eqnarray}
m_B^2&=& \Bigg[
56 e^{4}
-352 e^{3} g_{m}
+4\left(3+10 g_{l}+g_{l}^{2}\right) g_{m}^{4}
-4\left(29+2 g_{l}+29 g_{l}^{2}\right) g_{m}^{2}\lambda
\nonumber\\
&&
+\left(53+2 g_{l}+17 g_{l}^{2}\right)\lambda^{2}
-48 e^{2}\Big((-3+2 g_{l}) g_{m}^{2}+5\lambda\Big)
\nonumber\\
&&
+16 e\Big((8+6 g_{l}) g_{m}^{3}-15(1+g_{l}) g_{m}\lambda\Big)
\Bigg]\frac{\mu^{2}}{192\pi^{2}}.
\end{eqnarray}
The characterization of $B_c^2=\mu^2$ as a minimum of $V_{\rm eff}$ is therefore controlled by the sign of $m_B^2$. To delineate the regions of the parameter space $(e,\lambda,g_m,g_l)$ for which $m_B^2>0$, we fix two couplings and scan the remaining two-dimensional subspace. Figures~\ref{ps01}--\ref{ps03} display representative slices in the $(e,\lambda)$ plane, where the shaded region corresponds to $m_B^2(e,\lambda;g_m,g_l)>0$ for the fixed values of $(g_m,g_l)$ indicated in each panel.

In Fig.~\ref{ps01} we set $g_m=0$ and vary $g_l$ within the perturbative regime. We find that changing $g_l$ has only a mild impact on the shape of the allowed domain. In Fig.~\ref{ps02} we fix $g_l=0$\footnote{Since the consistency of the longitudinal sector requires $g_l>0$, the point $g_l=0$ is approached only as a boundary limit from the physical side of parameter space.} and vary $g_m$ over positive values, $0\le g_m\le 1$, while in Fig.~\ref{ps03} we again set $g_l=0$ and consider negative values, $-1\le g_m\le -0.01$. Altogether, these plots show that there exist perturbative regions of parameter space in which $m_B^2>0$, supporting a radiatively induced symmetry-breaking vacuum in the present model.

\begin{figure}[ht!]
	\includegraphics[angle=0,width=15cm]{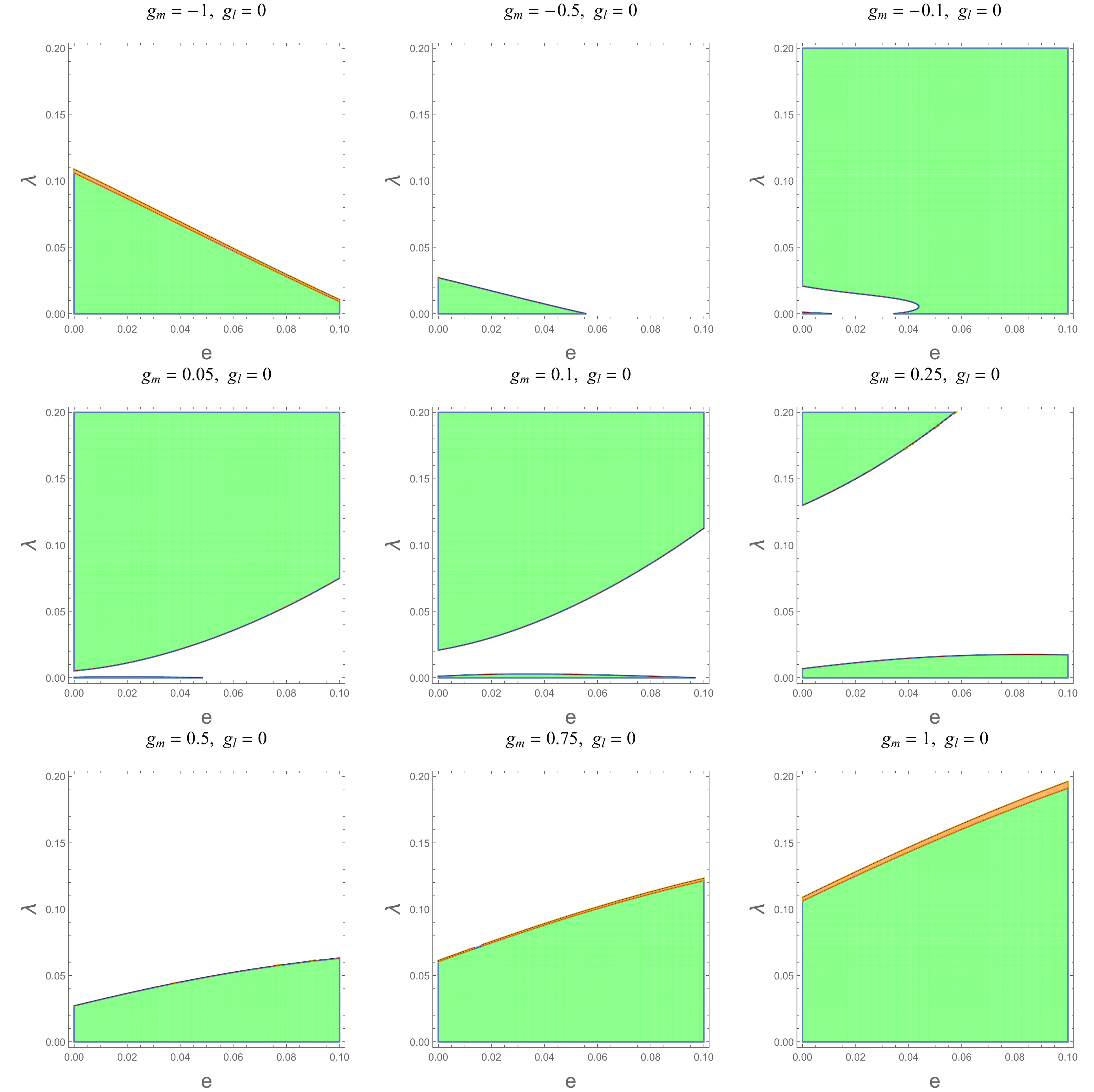}
	\caption{Comparison of the regions in the $(e,\lambda)$ plane where $m_B^2/\mu^2>0$, evaluated at lower LL order and after inclusion of the $L^4$ LL corrections, for representative values of $g_m$ and fixed $g_l=0$. The green region represents the subset of parameter space that remains positive in both approximations, whereas the orange region identifies points that are allowed at lower LL order but become excluded once the $L^4$ LL contributions are included. The black and dark-red curves mark the corresponding boundaries $m_B^2=0$ at lower LL order and at $L^4$ LL order, respectively. One observes that the higher-order LL terms induce only mild deformations of the positivity domain in the perturbative regime, although for $|g_m|\sim\mathcal{O}(1)$ a somewhat more visible excluded strip emerges.}
	\label{fig:l1l4comparisonpert}
\end{figure}

To further assess the robustness of the positivity condition $m_B^2>0$, we extended the LL analysis beyond the lowest nontrivial order and compared the region in the $(e,\lambda)$ plane defined by $m_B^2/\mu^2>0$ at lower LL order with the corresponding result including the $L^4$ LL contributions. The explicit expression obtained in this higher-order LL improvement is rather lengthy and was therefore omitted from the main text, since its full form does not provide additional conceptual insight. Instead, its effect is summarized in Fig.~\ref{fig:l1l4comparisonpert}, 
where we display the corresponding comparison for representative values of $g_m$ at fixed $g_l=0$. The main outcome is that, throughout the more perturbative regions of parameter space, the inclusion of the $L^4$ LL terms produces only very small deformations of the boundary of the domain where $m_B^2/\mu^2>0$. Even when the effect becomes more visible for larger magnetic-type coupling, particularly for $|g_m| \sim \mathcal{O}(1)$, the additional excluded portion of parameter space remains quantitatively narrow.

At the same time, this result should be interpreted with the appropriate perturbative caution. What the comparison shows is a local robustness of the positive-mass region within the truncated LL approximation considered here; it does not establish that the sign of $m_B^2$ is protected under the full renormalization-group flow at arbitrarily high energies. Since $m_B^2$ is determined by a specific combination of running couplings, the exact RG evolution may in principle drive the theory across the hypersurface $m_B^2=0$. Moreover, as usual in quantum field theory, the perturbative RG-improved expansion should be viewed as an asymptotic construction rather than as a globally convergent one. For this reason, our results support the statement that the radiatively induced vacuum is perturbatively stable in the weak-coupling regime probed here, but they should not be taken as a proof of all-orders or global sign preservation of the dynamically generated mass.

\section{Conclusions and outlook}
We investigated a complex extension of the bumblebee scenario coupled to an Abelian gauge field, focusing on its one-loop renormalization and on the radiative structure of its effective potential. The theory generalizes the usual real bumblebee dynamics by promoting the LV vector to a complex field and by allowing, besides the minimal coupling, a longitudinal kinetic term controlled by $g_l$ and a magnetic-type non-minimal interaction controlled by $g_m$. Among results of our paper, we can emphasize, first, formulation of a new coupling between gauge and bumblebee models, developing further ideas proposed in \cite{Lehum:2024ovo}, second, generalization of the bumblebee theory to the complex case. Then, working in dimensional regularization within the minimal-subtraction scheme, we determined the one-loop UV divergences of the Green functions required to renormalize the gauge, longitudinal, and quartic sectors. The corresponding counterterms yield the one-loop beta functions for $e$, $g_l$, $g_m$, and for the quartic self-couplings $\lambda$ and $\tilde\lambda$.

It is also worth emphasizing a structural consequence of the one-loop RG functions.
Once the gauge interaction is switched on, the hypersurface in coupling space defined by
$g_m=g_l=\lambda=\tilde\lambda=0$ is \emph{not} preserved by the RG flow:
even if these interactions are tuned to vanish at some reference scale, gauge fluctuations
radiatively regenerate them through genuine source terms in the beta functions,
cf.\ Eqs.~(\ref{Eq:betas}b)--(\ref{Eq:betas}e).
In particular, at $g_m=g_l=\lambda=\tilde\lambda=0$ one finds nonzero
$\beta(\lambda)\propto e^{4}$, $\beta(\tilde\lambda)\propto e^{4}$,
$\beta(g_m)\propto e^{3}$, and $\beta(g_l)\propto e^{2}$, see Eq.~(\ref{eq:betas_sources}).
From an effective-field-theory viewpoint, this reflects operator mixing among the marginal
interactions compatible with Lorentz symmetry and the $U(1)$ phase symmetry of $B_\mu$,
so that a consistent renormalization requires treating $(g_l,g_m,\lambda,\tilde\lambda)$ as
running couplings rather than as optional deformations.

We also developed an RG-covariant LL improvement scheme, motivated by the Vilkovisky--DeWitt formulation in normal field coordinates, and applied it to the constant real bumblebee background. This allowed us to obtain the corresponding one-loop LL effective potential and to identify perturbative regions of parameter space in which dimensional transmutation may trigger a nontrivial vacuum, providing a dynamical route to LSB in the present model. We expect that this methodology can be generalized to our LV models, in particular, those ones including gravity.

Several directions deserve further study. Extending the analysis beyond the LL approximation requires the two-loop renormalization-group functions and would enable a more stringent assessment of the vacuum structure. It is also of interest to investigate the spectrum of fluctuations about the radiatively induced minimum and to quantify the role of the longitudinal sector governed by $g_l$. Also, the notorious problem of the bumblebee model is related with its classical evolution, see the discussion in \cite{Bluhm:2004ep,Bluhm:2007bd}. While our theory essentially differs from the usual real bumblebee model due to the presence of the longitudinal term in the classical action, which certainly changes the algebra of Dirac constraints, a detailed study of this problem for our model certainly must be performed.
Finally, coupling the model to gravity~\cite{Lehum:2024ovo,Lehum:2024wmk} -- including possible non-minimal curvature interactions of the type considered in Ref.~\cite{Lehum:2024wmk} -- may help to elucidate the interplay between radiative Lorentz symmetry breaking and gravitational dynamics, and may further sharpen phenomenological constraints.

\vspace{.5cm}

\textbf{Acknowledgments.} 
This work was partially supported by Conselho Nacional de Desenvolvimento Cient\'{i}fico e Tecnol\'{o}gico (CNPq). W. C. is partially supported by Coordena\c{c}\~ao de Aperfei\c{c}oamento de Pessoal de N\'ivel Superior (CAPES). The work by A. Yu. P. has been partially supported by the CNPq project No. 303777/2023-0. The work of A. C. L. was partially supported by CNPq, Grants No.~404310/2023-0 and No.~301256/2025-0.

{\bf Data Availability Statement:} No Data associated in the manuscript.

{\bf Conflict of interest statement:} The authors declare that they have no conflict of interests.

 \appendix

\section*{APPENDIX A: BASIC INTEGRALS}\label{app:pvi}

We work in dimensional regularization with $D=4-2\epsilon$, with the renormalization scale $\mu$. Following the standard FeynCalc \cite{feyncalc} normalization, the scalar one-loop integrals are defined by
\begin{equation}
\frac{(2\pi\mu)^{4-D}}{i\pi^{2}}
\int d^{D}q \; \frac{1}{\prod_{j}\big[(q+r_j)^2-m_j^2\big]}\,,
\end{equation}
where the shifts $r_j$ encode the external kinematics. In particular, the scalar Passarino--Veltman functions used throughout this work are
\begin{eqnarray}
\textbf{A}_0(m^2)
&=&
\frac{(2\pi\mu)^{4-D}}{i\pi^{2}}
\int d^{D}q \;\frac{1}{q^{2}-m^{2}}=\frac{m^2}{\epsilon}+\text{finite}\,,
\label{eq:A0def}
\\
\textbf{B}_0(p^2;m_1^2,m_2^2)
&=&
\frac{(2\pi\mu)^{4-D}}{i\pi^{2}}
\int d^{D}q \;\frac{1}{\big(q^{2}-m_1^{2}\big)\big((q+p)^{2}-m_2^{2}\big)}=\frac{1}{\epsilon}+\text{finite}\,,
\label{eq:B0def}
\\
\textbf{C}_0(p_i^2;m_i^2)
&=&
\frac{(2\pi\mu)^{4-D}}{i\pi^{2}}
\int d^{D}q~\frac{1}{\big(q^{2}-m_1^{2}\big)\big((q+p_1)^{2}-m_2^{2}\big)}\nonumber\\
&&\frac{1}{\big((q+p_1+p_2)^{2}-m_3^{2}\big)}=\text{finite}\,,
\label{eq:C0def}
\\
\textbf{D}_0(p_i^2;m_i^2)
&=&
\frac{(2\pi\mu)^{4-D}}{i\pi^{2}}
\int d^{D}q \;\frac{1}{\big(q^{2}-m_1^{2}\big)\big((q+p_1)^{2}-m_2^{2}\big)}\nonumber\\
&&\times\frac{1}{\big((q+p_1+p_2)^{2}-m_3^{2}\big)\big((q+p_1+p_2+p_3)^{2}-m_4^{2}\big)}=\text{finite}\,.
\label{eq:D0def}
\end{eqnarray}

\section*{APPENDIX B: CHANGE OF VARIABLES TO RG--COVARIANT COORDINATES AND CANCELLATION OF THE ANOMALOUS-DIMENSION TERM}\label{sec:change-of-vars}

We consider a renormalized functional $F[\phi,\lambda,\mu]$ (e.g. the effective action)
of renormalized fields $\phi^A(x)$, renormalized couplings $\lambda_i$, and scale $\mu$.
Introduce the RG--covariant field coordinates $\chi^A$ via a nonsingular map
$\phi=\phi(\chi,\lambda,\mu)$ (in the VDW construction this is the exponential map on
field space, i.e. normal geodesic coordinates). Define
\begin{equation}
\tilde F(\chi,\lambda,\mu)\;\equiv\;F\big(\phi(\chi,\lambda,\mu),\lambda,\mu\big).
\end{equation}
Our goal is to transform the anomalous--dimension term
\begin{equation}
\label{eq:I-def}
\mathcal{I}\;\equiv\;\int d^Dx\;\gamma^{\ A}{}_{B}\,\phi^B(x)\,\frac{\delta F}{\delta\phi^A(x)}
\end{equation}
to $\chi$--variables and show how it cancels in the RG operator taken at fixed $\chi$.

By the functional chain rule,
\begin{equation}
\label{eq:chain}
\frac{\delta F}{\delta\phi^A(x)}
\;=\;\frac{\partial\tilde F}{\partial\chi^C}\,
\frac{\delta\chi^C}{\delta\phi^A(x)}\,,
\end{equation}
so that \eqref{eq:I-def} becomes
\begin{equation}
\label{eq:I-K}
\mathcal{I}
\;=\;\int d^Dx\;\gamma^{\ A}{}_{B}\,\phi^B(x)\,
\frac{\delta\chi^C}{\delta\phi^A(x)}\;\frac{\partial\tilde F}{\partial\chi^C}
\;\equiv\;\mathcal{K}^C\,\frac{\partial\tilde F}{\partial\chi^C},
\qquad
\mathcal{K}^C\;\equiv\;\int d^Dx\;\gamma^{\ A}{}_{B}\,\phi^B\,
\frac{\delta\chi^C}{\delta\phi^A}.
\end{equation}

"Fixed $\chi$" means by definition
\begin{equation}
\label{eq:chi-fixed}
\Big(\mu\tfrac{d}{d\mu}\Big)_{\!\chi}\chi^C(x)=0
\quad\Rightarrow\quad
\mu\frac{\partial\chi^C}{\partial\mu}
+\sum_i\beta_i\,\frac{\partial\chi^C}{\partial\lambda_i}
+\int d^Dy\;\frac{\delta\chi^C}{\delta\phi^A(y)}
\Big(\mu\tfrac{d\phi^A(y)}{d\mu}\Big)_{\!\chi}=0.
\end{equation}
At fixed $\chi$ the renormalized field runs according to its anomalous dimension,
\begin{equation}
\label{eq:phi-flow}
\Big(\mu\tfrac{d\phi^A}{d\mu}\Big)_{\!\chi}=-\,\gamma^{\ A}{}_{B}\,\phi^B.
\end{equation}
Inserting \eqref{eq:phi-flow} into \eqref{eq:chi-fixed} yields the \emph{key identity}
\begin{equation}
\label{eq:K-identity}
\mathcal{K}^C
=\mu\,\frac{\partial\chi^C}{\partial\mu}
+\sum_i\beta_i\,\frac{\partial\chi^C}{\partial\lambda_i}.
\end{equation}

Consider the total $\mu$--derivative at fixed $\chi$ acting on $F$,
\begin{equation}
\label{eq:RG-op-phi}
\Big(\mu\tfrac{d}{d\mu}\Big)_{\!\chi}F
=\mu\partial_\mu F+\sum_i\beta_i\,\partial_{\lambda_i}F
-\int d^Dx\;\gamma^{\ A}{}_{B}\,\phi^B\,\frac{\delta F}{\delta\phi^A}.
\end{equation}
Change variables $(\phi\to\chi)$ in the explicit derivatives:
\begin{equation}
\label{eq:explicit-derivs}
\mu\partial_\mu F=\mu\partial_\mu\tilde F
+\Big(\mu\,\partial_\mu\chi^C\Big)\frac{\partial\tilde F}{\partial\chi^C},
\qquad
\partial_{\lambda_i}F=\partial_{\lambda_i}\tilde F
+\Big(\partial_{\lambda_i}\chi^C\Big)\frac{\partial\tilde F}{\partial\chi^C}.
\end{equation}
Using \eqref{eq:I-K} and \eqref{eq:K-identity} one has
\begin{equation}
\label{eq:I-final}
\mathcal{I}
=\Big(\mu\,\partial_\mu\chi^C+\sum_i\beta_i\,\partial_{\lambda_i}\chi^C\Big)
\frac{\partial\tilde F}{\partial\chi^C}.
\end{equation}
Substituting \eqref{eq:explicit-derivs} and \eqref{eq:I-final} into
\eqref{eq:RG-op-phi}, the cross terms cancel \emph{exactly}:
\begin{equation}
\label{eq:RG-op-chi}
\Big(\mu\tfrac{d}{d\mu}\Big)_{\!\chi}F\big(\phi(\chi),\lambda,\mu\big)
=\mu\partial_\mu\tilde F(\chi,\lambda,\mu)
+\sum_i\beta_i\,\partial_{\lambda_i}\tilde F(\chi,\lambda,\mu).
\end{equation}
Thus, when the RG operator is taken at fixed RG--covariant field $\chi$, the
anomalous--dimension term in the $\phi$--representation is absorbed by the
implicit $(\mu,\lambda)$--dependence of $\chi$.

\section*{APPENDIX C: SCALAR QED AS AN STANDARD EXAMPLE}\label{example_sqed}

To illustrate the general procedure of introducing the RG-covariant coordinates, we apply it to scalar QED and compute the LL VDW effective potential. The classically scale-invariant Lagrangian density reads
\begin{eqnarray}
\mathcal{L}
&=& -\frac{1}{2} F^{\mu\nu}F_{\mu\nu}
+ (D^\mu\phi)^*\,D_\mu\phi
- \frac{\lambda}{4!}\,(\phi^*\phi)^2,
\end{eqnarray}
where $F_{\mu\nu} = \partial_\mu A_\nu - \partial_\nu A_\mu$ is the Maxwell field strength, $D_\mu\phi = (\partial_\mu - i e\,A_\mu)\phi$ is the covariant derivative, $\phi$ is a complex scalar field, and $A_\mu$ denotes the $U(1)$ gauge field.

From the geometric viewpoint, the RG-covariant fields introduced in Sec.~\ref{sec:RGcov} are defined via the exponential map on field space, so that the original fields are obtained as
\begin{equation}
\phi^A(\chi) = \exp_{\bar\phi}(\chi)^A,
\end{equation}
along geodesics emanating from a reference configuration $\bar\phi^A$. Expanding this relation in powers of the geodesic coordinate $\chi^A$ yields the standard normal-coordinate series
\begin{equation}
\phi^A(\chi)
= \bar\phi^A
+ \chi^A
- \frac{1}{2}\,\bar\Gamma^{A}{}_{BC}(\bar\phi)\,\chi^B\chi^C
+ \mathcal{O}(\chi^3),
\label{eq:normal-expansion}
\end{equation}
where $\bar\Gamma^{A}{}_{BC}(\bar\phi)$ is the connection evaluated at the base point. By definition of Riemann normal coordinates, one chooses $\chi^A$ so that $\bar\Gamma^{A}{}_{BC}(\bar\phi)=0$, implying that the leading nonlinearities in~\eqref{eq:normal-expansion} arise at order $\chi^3$ and are controlled by the curvature of the field manifold. In a generic curved field space this construction is inherently local, providing a locally linear parametrization only in a neighborhood of $\bar\phi^A$.

In the scalar models considered here, however, the physical orbit space relevant for the VDW potential is one-dimensional and flat. For a single real scalar with canonical kinetic term the DeWitt metric is trivial, so the background field itself is a global normal coordinate. For a complex scalar with potential $V_0(\phi^*\phi)=\lambda(\phi^*\phi)^2/4!$ (with or without the $U(1)$ gauge interactions of scalar QED), the physical configuration space after quotienting by the $U(1)$ orbits is parametrized by the radial modulus $\rho=(\phi^*\phi)^{1/2}$, and the induced line element is $ds^2=d\rho^2$. In this one-dimensional flat manifold the connection vanishes identically, $\Gamma^\rho{}_{\rho\rho}=0$, so the exponential map reduces to a global affine relation and $\rho$ itself is a Riemann normal coordinate. Consequently, within the perturbative domain of the theory we can simply identify the RG-covariant field with this radial coordinate, $\chi=\rho$, or equivalently
\begin{equation}
\chi^2 \equiv \phi^*\phi,
\end{equation}
and formulate the VDW potential and its RG improvement directly in terms of the single normal, RG-invariant field $\chi$. Although, in general, the base point $\bar\phi$ for the normal-coordinate construction can be chosen arbitrarily in field space, in vacuum applications it is natural to take $\bar\phi$ to be the vacuum configuration, i.e., the minimum of the effective potential. In our scalar-QED example, however, the one-dimensional flat orbit space implies that the radial modulus $\chi$ serves as a global Riemann normal coordinate, essentially independent of the specific choice of reference point, as long as the perturbative description remains valid.

With this choice of background variable, the RG input to the LL resummation is provided by the one-loop beta functions, which for scalar QED are well known~\cite{Peskin:1995ev,Srednicki:2007qs},
\begin{eqnarray}
\beta_\lambda &=& \frac{5\lambda^2 - 18\lambda\,e^2 + 54 e^4}{24\pi^2},\nonumber\\
\beta_e &=& \frac{e^3}{48\pi^2}.
\end{eqnarray}

The LL VDW effective potential is then constructed from the ansatz~\eqref{eq:ansatz1}, with the LL coefficients obeying the general recursion relation~\eqref{recursive_relation1}. In the present case, this relation takes the explicit form
\begin{eqnarray}\label{recursive_relation_sqed}
C_{LL}^{(n)}(\lambda,e)
= \frac{1}{2n}
\left(
\beta_\lambda\frac{\partial}{\partial \lambda}
+ \beta_e\frac{\partial}{\partial e}
\right) C_{LL}^{(n-1)}(\lambda,e),
\qquad n \ge 1,
\end{eqnarray}
with $C_{LL}^{(0)}(\lambda,e) = \lambda/24$. Up to order $L^4$, the LL coefficients of the VDW effective potential are
\begin{subequations}
\begin{eqnarray}
C_{LL}^{(1)} &=& \frac{54 e^4 - 18 e^2 \lambda + 5 \lambda^2}{1152 \pi^2}, \\
C_{LL}^{(2)} &=& \frac{-432 e^6 + 423 e^4 \lambda - 135 e^2 \lambda^2 + 25 \lambda^3}{55296 \pi^4}, \\
C_{LL}^{(3)} &=& \frac{7182 e^8 - 7116 e^6 \lambda + 3630 e^4 \lambda^2 - 900 e^2 \lambda^3 + 125 \lambda^4}{2654208 \pi^6}, \\
C_{LL}^{(4)} &=& \frac{-88884 e^{10} + 124695 e^8 \lambda - 76200 e^6 \lambda^2 + 27750 e^4 \lambda^3 - 5625 e^2 \lambda^4 + 625 \lambda^5}{127401984 \pi^8}.
\end{eqnarray}
\end{subequations}
It is important to note that, for power-counting purposes, we treat both $e^2$ and $\lambda$ as quantities of order $x$, i.e. $e^2 \sim x$ and $\lambda \sim x$.

The CW renormalization condition~\eqref{cw:ren_cond} fixes the quartic counterterm to be
\begin{eqnarray}\label{delta_lambda}
\delta_\lambda &=& \frac{1}{331776 \pi^8}\Bigg[
88884 e^{10}
- 45 e^8 \left(2771 \lambda + 9576 \pi^2\right)
+ 120 e^6 \left(635 \lambda^2 + 3558 \pi^2 \lambda + 6048 \pi^4\right)\nonumber\\
&&\qquad\qquad\qquad
- 30 e^4 \left(925 \lambda^3 + 7260 \pi^2 \lambda^2 + 23688 \pi^4 \lambda + 51840 \pi^6\right)\nonumber\\
&&\qquad\qquad\qquad
+ 225 e^2 \lambda \left(25 \lambda^3 + 240 \pi^2 \lambda^2 + 1008 \pi^4 \lambda + 2304 \pi^6\right)\nonumber\\
&&\qquad\qquad\qquad
- 125 \lambda^2 \left(5 \lambda^3 + 60 \pi^2 \lambda^2 + 336 \pi^4 \lambda + 1152 \pi^6\right)
\Bigg].
\end{eqnarray}

Inserting Eq.~\eqref{delta_lambda} into the LL VDW effective potential, we obtain
\begin{eqnarray}
V_{_{\rm VDW}}^{LL}
&=& \chi^4 \left(\sum_{n=0}^{4} A_n\,L^n\right),
\end{eqnarray}
with
\begin{subequations}
\begin{eqnarray}
A_0 &=& -\frac{625 \lambda^5}{7962624 \pi^8}
+ \frac{\left(5625 e^2 - 7500 \pi^2\right) \lambda^4}{7962624 \pi^8}
+ \frac{\left(-27750 e^4 + 54000 \pi^2 e^2 - 42000 \pi^4\right) \lambda^3}{7962624 \pi^8}\nonumber\\
&&+\frac{\left(76200 e^6 - 217800 \pi^2 e^4 + 226800 \pi^4 e^2 - 144000 \pi^6\right) \lambda^2}{7962624 \pi^8}\nonumber\\
&&+\frac{\left(-124695 e^8 + 426960 \pi^2 e^6 - 710640 \pi^4 e^4 + 518400 \pi^6 e^2 + 331776 \pi^8\right) \lambda}{7962624 \pi^8}\nonumber\\
&&+\frac{88884 e^{10} - 430920 \pi^2 e^8 + 725760 \pi^4 e^6 - 1555200 \pi^6 e^4}{7962624 \pi^8},\\[4pt]
A_1 &=& \frac{5 \lambda^2 - 18 e^2 \lambda + 54 e^4}{1152 \pi^2},\\[4pt]
A_2 &=& \frac{25 \lambda^3 - 135 e^2 \lambda^2 + 423 e^4 \lambda - 432 e^6}{55296 \pi^4},\\[4pt]
A_3 &=& \frac{125 \lambda^4 - 900 e^2 \lambda^3 + 3630 e^4 \lambda^2 - 7116 e^6 \lambda + 7182 e^8}{2654208 \pi^6},\\[4pt]
A_4 &=& \frac{625 \lambda^5 - 5625 e^2 \lambda^4 + 27750 e^4 \lambda^3 - 76200 e^6 \lambda^2 + 124695 e^8 \lambda - 88884 e^{10}}{127401984 \pi^8}.
\end{eqnarray}
\end{subequations}

The location of the minimum follows from the condition~\eqref{cond_gap}, which can be solved perturbatively for $\lambda$ in powers of $e$:
\begin{eqnarray}
\lambda
&=& \frac{33\, e^4}{8 \pi^2}
- \frac{503\, e^6}{64 \pi^4}
+ \frac{56197\, e^8}{2048 \pi^6}
- \frac{4745009\, e^{10}}{49152 \pi^8}
+ \mathcal{O}(e^{12}).
\end{eqnarray}
Substituting this expression into $V_{_{\rm VDW}}^{LL}$, we arrive at
\begin{eqnarray}
V_{_{\rm VDW}}^{LL}
&=& -\frac{e^4 \chi^4}{2359296 \pi^8}
\Bigg[
55296 \pi^6
- 76032 \pi^4 e^2
+ 231984 \pi^2 e^4
- 837753 e^6 \nonumber\\
&&\quad
+ 6 \left(279251 e^6 - 77328 \pi^2 e^4 + 25344 \pi^4 e^2 - 18432 \pi^6\right)
\ln\!\left(\frac{\chi^2}{\mu^2}\right)\nonumber\\
&&\quad
+ 48 \left(4997 e^6 - 1551 \pi^2 e^4 + 384 \pi^4 e^2\right)
\ln^2\!\left(\frac{\chi^2}{\mu^2}\right)\nonumber\\
&&\quad
+ 4 \left(6523 e^6 - 1596 \pi^2 e^4\right)
\ln^3\!\left(\frac{\chi^2}{\mu^2}\right)
+ 1646 e^6 \ln^4\!\left(\frac{\chi^2}{\mu^2}\right)
\Bigg].
\end{eqnarray}

The dynamically generated mass of the scalar mode is obtained from the curvature at the minimum, Eq.~\eqref{cond_mass}, which yields
\begin{eqnarray}
m^2_{\chi}
&=& \frac{e^4\,\mu^2}{8\pi^2}
\left[
3
- \frac{37\, e^2}{8 \pi^2}
+ \frac{3739\, e^4}{256 \pi^4}
- \frac{106409\, e^{6}}{2048 \pi^6}
\right]
+ \mathcal{O}(e^{12}) > 0.
\end{eqnarray}

Thus, the VDW framework, combined with the RG-improved LL resummation, sharpens the determination of the radiatively generated mass starting from one-loop beta functions, while maintaining manifest gauge and parametrization independence of the effective potential throughout.



\end{document}